\documentclass[showpacs,aps,pra,twocolumn, groupedaddress]{revtex4}
\usepackage{array}
\usepackage{times,amsmath,graphicx,amssymb}
\usepackage{color}
\usepackage{ulem}
\usepackage{soul}

\usepackage{tikz}
\usetikzlibrary{matrix}

\definecolor{grey}{rgb}{.6,.6,.6}
\definecolor{forestgreen}{rgb}{.13,.55,.13}
\definecolor{amber}{rgb}{1.00,.31,.0}

\def \hH{ \hat{\mathcal{H}}}
\def \bse{\begin{subequations}}
\def \ese{\end{subequations}}

\def \vwg{v_{\rm G}}

\newcommand{\PD}{\phantom{\dag}}
\newcommand{\ev}[1]{\ensuremath{\left\langle #1 \right\rangle}}

\graphicspath{{./figures/}}

\begin{document}

\title{Nonreciprocal Photon Transmission and Amplification via Reservoir Engineering}

\author{A. Metelmann}
\author{A. A. Clerk}
\affiliation{Department of Physics, McGill University, 3600 rue University, Montr\'{e}al, Quebec, H3A 2T8 Canada}
\date{\today}

\begin{abstract}
We discuss a general method for constructing nonreciprocal, cavity-based photonic devices, based on matching a given
coherent interaction with its corresponding dissipative counterpart; our method generalizes the basic structure used in the theory
of cascaded quantum systems,  and can render an extremely wide class of interactions directional.  In contrast to standard interference-based schemes, our approach allows directional behavior over a wide bandwidth. 
We show how it can be used to devise isolators and directional, quantum-limited amplifiers.  We discuss in detail how 
this general method allows the construction of a directional, noise-free phase-sensitive amplifier that is not limited by any fundamental
gain-bandwidth constraint.   Our approach is particularly well-suited to implementations using superconducting microwave circuits
 and optomechanical systems.
\end{abstract}
	
\pacs{42.65.Yj, 03.65.Ta, 84.30.Le}



\maketitle


\section{Introduction}

The general desire to break time-reversal symmetry and reciprocity in engineered photonic structures has garnered an immense amount of
recent interest.  Recall that while time-reversal symmetry is only a useful notion in non-dissipative systems, reciprocity is more general:  
it is defined as the invariance of photon transmission amplitudes under exchange of source and detector \cite{Deak2012}.  On a fundamental level, the artificial breaking of time-reversal symmetry allows the realization of truly new photonic states, such as quantum Hall states and more general topological states \cite{Raghu2008,Haldane2008,Wang2008,Hafezi2011,Umacalilar2011}.  
On a more practical level, nonreciprocal devices can enable a number of signal-processing applications 
and greatly simplify the construction of photonic networks \cite{Jalas2013}.  

Nonreciprocal microwave-frequency devices are also crucial to efforts at quantum-information processing with superconducting circuits.  Here one necessarily needs to use near quantum-limited amplifiers to efficiently read out qubits; nonreciprocity is crucial to ensure the qubits are protected from unwanted noise stemming from the amplifier.  The conventional solution is to use circulators employing magneto-optical effects (Faraday rotation)
to break reciprocity.  These devices have many disadvantages:  they are bulky and cannot be implemented on-chip (hindering scaling-up to multi-qubit systems), and they use large magnetic fields, which can be deleterious to superconducting devices.  Their use also typically leads to insertion losses.

A number of strategies have been developed to break reciprocity without the use of magneto-optical effects in both optical systems and superconducting circuits.  For nonreciprocal photon transmission, approaches based on refractive-index modulation \cite{Yu2009,Lira2012} have been considered, as well as strategies using optical non-linearity \cite{Soljacic2003}, optomechanical interaction \cite{Manipatruni2009,Hafezi2012}, and interfering parametric processes \cite{Kamal2011,Kamal2014, Kerckhoff2015}.  Related strategies where the phases of external driving fields generate an artificial gauge field in a lattice or cavity array have also been discussed \cite{Fang2012,Fang2012b,Tzuang2014,Estep2014}, as have alternative methods that do not use modulation \cite{Koch2010,Nunnenkamp2011, Viola2014}.
Nonreciprocal quantum amplifiers have also been developed largely in the context of superconducting circuits \cite{Abdo2013b, Abdo2014,Ranzani2014a,Ranzani2014}.  They typically involve engineering complex interferences between parametric processes. Understanding how to achieve such interferences can be difficult, though recently a graph-theory approach was formulated by Ranzani et al. \cite{Ranzani2014a}.

 \begin{figure} 
 \centering\includegraphics[width=0.45\textwidth]{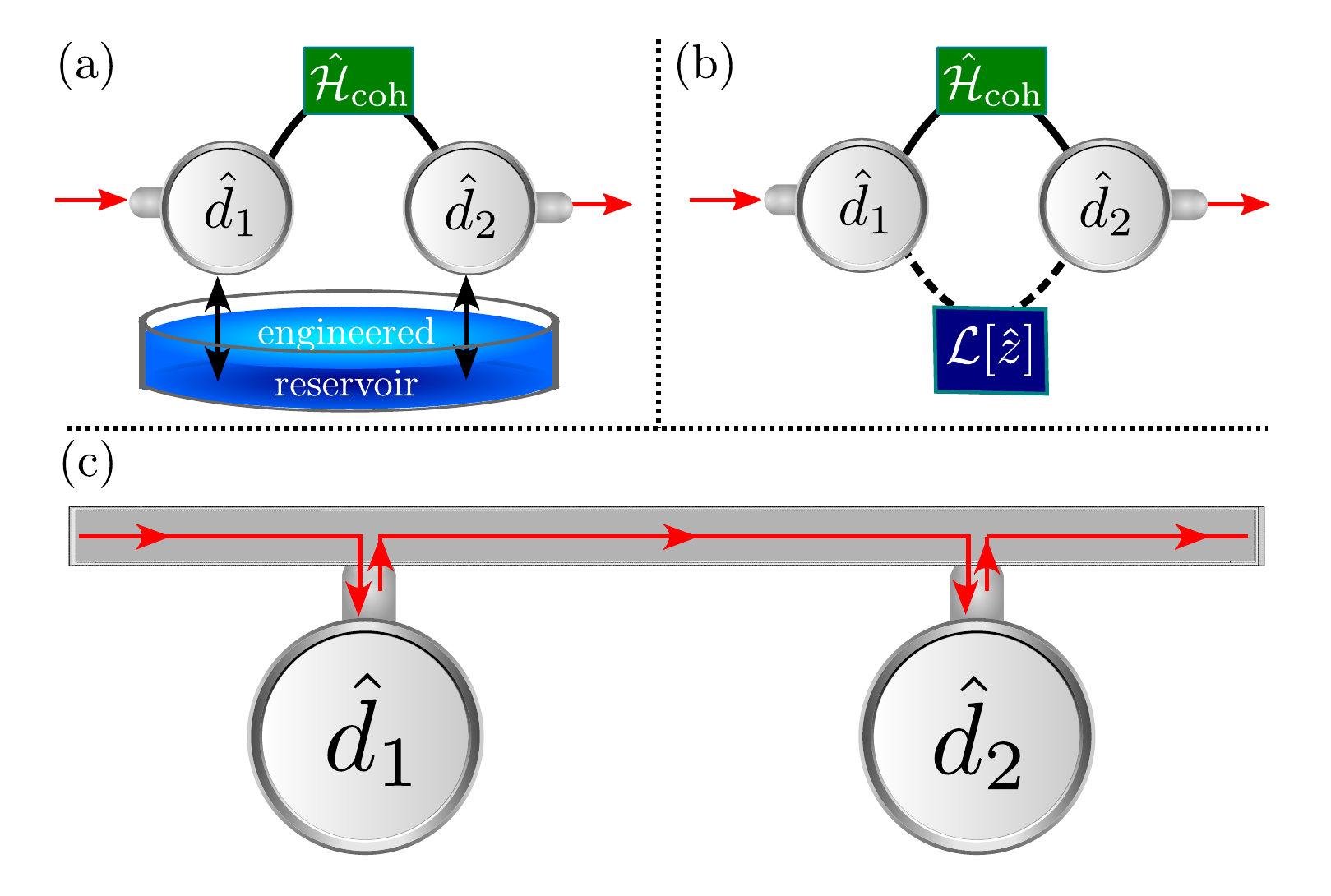} 
	\caption{
	 	(a) Basic recipe for generating directionality: two cavities are directly coupled to one another via a coherent Hamiltonian
		$\hH_{\rm coh}$, and are also each coupled to the same (non-directional) dissipative environment.
		(b) The dissipative environment in (a) mediates a reciprocal dissipative interaction between the two cavities.   This can be modeled
		using a Lindblad master equation and dissipative superoperator $\mathcal{L}[\hat{z}]$ (cf.~Eq.~(\ref{Eq.:GeneralMaster})).
		By balancing the strength of coherent and bath-induced dissipative interactions between the cavities, one can break reciprocity.
		(c) Schematic cascaded quantum system, where one cavity drives another via a waveguide supporting only
		a right-propagating mode. The effective theory used to describe such systems corresponds to (b).}
	\label{fig:SketchCascaded}
 \end{figure}

In this work, we present a simple yet general method for generating nonreciprocal behavior in a photonic system, 
one that can make a variety of cavity-cavity interactions completely directional, including amplifying interactions 
(see Fig.~\ref{fig:SketchCascaded}).  It employs reservoir engineering \cite{Poyatos1996},
where a structured dissipative environment generates useful quantum behavior.  In our approach,  
the dissipative reservoir (which could simply be a damped auxiliary cavity mode) generates an effective dissipative interaction between the modes of interest.  Nonreciprocal behavior is then obtained by balancing this induced dissipative interaction against the corresponding coherent version of the interaction.  

As we discuss, this simple yet powerful trick allows one to generate both isolators (which only allow unidirectional transmission), and nonreciprocal, quantum-limited phase-sensitive amplifiers (which have zero added noise) as well as phase-preserving amplifiers (which add the quantum-limited amount of noise, a half quantum at the signal frequency). While our approach uses a kind of interference, it is markedly different from more typical interference-based approaches, in that it allows perfect directional behavior over a wide range of frequencies. The method is also simple enough that it could be implemented in a wide variety of architectures; in particular, it is extremely well suited to implementations using superconducting circuits  and optomechanics. 

Our approach to nonreciprocity is intimately connected to the theory of cascaded quantum systems \cite{Carmichael1993, Gardiner1993}.  This is an effective theory developed to describe situations where a nonreciprocal element is used to couple two quantum systems (e.g.,~such that the output field of one cavity drives a second cavity, but  not vice-versa, see Fig.~\ref{fig:SketchCascaded}(c)).  We show that the effective interactions used in this theory have {\it exactly} the form described above:  one balances a coherent ``photon tunneling" interaction between the two cavities against a corresponding dissipative version of this interaction.  We also demonstrate that cascaded quantum systems theory is not simply an effective theory for describing nonreciprocal transmission: it also serves as a recipe for constructing nonreciprocal devices, one that can be generalized to amplifying interactions.  As we discuss, the needed dissipative interactions can be obtained by simply coupling to intermediate 
damped cavity modes; one does not need to start with an explicitly nonreciprocal reservoir (as assumed in the derivations of Refs.~\cite{Carmichael1993, Gardiner1993}).

The remainder of this paper is organized as follows.  In Sec.~\ref{Sec:Two:DirectionalityApplications}, we introduce our basic approach of balancing coherent and dissipation interactions, showing how this can be used to generate both nonreciprocal photon transmission as well as amplification.  
In Sec.~\ref{Sec.:Three:Implementations}, we provide further details on each of these schemes, discussing simple 3-mode implementations, 
as well as issues of bandwidth, impedance matching and added noise.  We pay particular attention to our scheme for a nonreciprocal cavity-based phase-sensitive amplifier.  In addition to being nonreciprocal and quantum-limited, we show that this system can also be constructed so that there is no fundamental gain-bandwidth limitation on its performance, and so that it is perfectly impedance matched at both its input and output (i.e.,~there are no unwanted reflections at either port of the amplifier).


\section{Directionality from dissipative interactions}\label{Sec:Two:DirectionalityApplications}

Throughout this work, we consider a generic situation where we have a pair of cavity modes (annihilation operators $\hat{d}_1,\hat{d}_2$), each coupled to input/output waveguides; our goal is to engineer a nonreciprocal interaction between them, thus enabling either nonreciprocal transmission or amplification of signals incident on the two modes.  Our approach is sketched in Fig.~\ref{fig:SketchCascaded}(b):  we allow both cavities to interact with one another in two distinct ways.  The first is via a direct, coherent interaction described by an interaction Hamiltonian  $\hH_{\rm coh}$.
While our approach can make a general factorizable cavity-cavity interaction directional, we focus here on simple bilinear interactions. 
The coherent interaction will thus be described by a quadratic Hamiltonian, having the general form  ($\hbar=1$) 
\begin{equation}
	\hH_{\rm coh} = J \hat{d}_1^\dagger \hat{d}_2 + \lambda \hat{d}_1^\dagger \hat{d}_2^\dagger + h.c. ,
	\label{Eq:Hint}
\end{equation}
where $J$ and $\lambda$ are in general complex.  We always work in a rotating frame where the two cavities are effectively resonant, and where $\hH_{\rm coh}$ is time-independent. Each of the two interactions in this Hamiltonian could be realized in many ways; for example, one could start with three modes and a generic
three-wave mixing Hamiltonian, and then displace one of the modes with a coherent tone.  The driven modes act as pumps; by a suitable choice of frequencies
 (i.e.,~at the difference and the sum of cavities 1 and 2 resonance frequencies), one realizes 
the above Hamiltonian, with the amplitudes and phases of the couplings $J, \lambda$ being controlled by the pump modes amplitudes.  Such an approach has been exploited recently in superconducting circuits, using the Josephson parametric converter (JPC) geometry \cite{Bergeal2010a,Bergeal2010,Abdo2011,Abdo2013}, as well as in quantum optomechanics \cite{Aspelmeyer2014}.

The second required interaction involves controllably coupling both cavities to the same dissipative reservoir (Fig.~\ref{fig:SketchCascaded}(a)).  Eliminating this reservoir will generate an effective dissipative interaction between the cavities (i.e.,~one that cannot be described by some direct Hamiltonian coupling). 
 The simplest setting is where this reservoir is effectively Markovian, and hence can be described using dissipators in a Lindblad master equation for the reduced density matrix $\hat{\rho}$ of the two cavity modes.  
 As we are focusing here on a bilinear coherent interaction, the needed interactions between the engineered reservoir and the two cavities 
will also be linear.  We are thus left with the general master equation   
\begin{align}\label{Eq.:GeneralMaster}
	 \frac{d}{dt} \hat \rho =& - i   \left[\hH_{\rm coh} , \hat \rho \right] 
			 + \Gamma \mathcal L [\hat z ] \hat{\rho}
			 +    \sum_{j = 1,2}  \kappa_j   \mathcal L [\hat d_j ] \hat{\rho},  
\end{align}
where
\begin{align}
	\hat{z} =& \sum_{j=1,2} \left(
		u_j \hat{d}_j + v_j \hat{d}_j^{\dagger}
	\right),
	\label{Eq:zDefn}
\end{align}
and the standard dissipative superoperator $\mathcal{L}[\hat{o}]$ is defined as
\begin{align}
	\mathcal L [\hat o ] \hat{\rho} =& \hat o  \hat \rho  \hat o^{\dag} 
					- \frac{1}{2} \hat o^{\dag} \hat o \hat \rho - \frac{1}{2} \hat \rho \hat o^{\dag} \hat o.  
\end{align}

The first term in Eq.~(\ref{Eq.:GeneralMaster}) describes the coherent interaction between the two cavities, the second the interaction with the engineered reservoir at rate $\Gamma$ (including the induced dissipative cavity-cavity interactions), and the last the coupling of the cavities to their input-output ports at rate  $\kappa_j $ .  Note that an asymmetry in the couplings does not change the basic physics in which we are interested; thus, for simplicity, we take $\kappa_1 = \kappa_2 \equiv \kappa$ in what follows. 
The coefficients $u_j$ and $v_j$ characterize the individual coupling of each cavity to the engineered bath.
As we see in what follows, the engineered reservoir need not be anything too exotic:  it can simply be another (damped) cavity mode, or a (non-directional) transmission line. Also note that one does not need to be in the strict Markovian limit, though it makes it simpler to understand the physics.  We discuss corrections to the Markovian limit in Sec.~\ref{Sec.:Three:Implementations}.

With these ingredients in place, obtaining directionality involves first constructing $\hH_{\rm coh}$ so that it gives the desired behavior (amplification or transmission), and then precisely balancing it with the corresponding dissipative interaction (i.e.,~choice of $\Gamma, u_j$ and $v_j$).  
To illustrate this, we can derive the equations of motion for the expectation values of the mode's operators. Starting from the Lindblad master equation in Eq.~(\ref{Eq.:GeneralMaster}) we obtain  
\begin{align} 
	 \frac{d}{dt} \ev{\hat d_1} =&     
 		- \frac{\Gamma_1 + \kappa }{2} \ev{ \hat d_1}  
\nonumber \\ &
 		- i \left[ J    \ \   + i \mu  \    \frac{\Gamma}{2}  \right]    \ev{ \hat d_2 }   
		- i \left[ \lambda   + i  \nu \frac{\Gamma}{2}  \right]    \ev{  \hat d_2^{\dag} } , 
\nonumber \\
	 \frac{d}{dt} \ev{\hat d_2} =&     
 		- \frac{\Gamma_2 + \kappa}{2}  \ev{ \hat d_2}   
\nonumber \\ &           
 		- i \left[J^{\ast} + i  \mu^{\ast} \frac{ \Gamma}{2} \right]   \ev{\hat d_1 }
 		- i \left[ \lambda - i  \nu  \frac{ \Gamma}{2}  \right]    \ev{ \hat d_1^{\dag}} ,  
\end{align} 
with $\Gamma_n = \Gamma ( |u_n|^2 -|v_n|^2 ), \ (n \in 1,2) $ describing the local damping induced by the engineered reservoir, and the definitions
$\mu = v_1 v_2^{\ast} - u_2  u_1^{\ast}$ and $ \nu = v_1 u_2^{\ast} - v_2  u_1^{\ast} $. The engineered reservoir mediates a non-local damping force on each mode, thus it couples the two modes in a similar manner as the coherent interaction. Crucially,  due to the difference in the coupling coefficients  we can decouple
 cavity  $1$ from cavity $2$ by setting 
\begin{align}
 J    \ \   \overset{!}{=} - i \mu  \    \frac{\Gamma}{2}, 
\hspace{0.5cm}
 \lambda    \overset{!}{=} - i  \nu \frac{\Gamma}{2}.
\end{align}
For this case we obtain a uni-directional interaction, where cavity $2$ is driven by cavity $1$ but not vice versa.
Moreover, it is straightforward to show that this decoupling works for all operators:  the evolution of any cavity-1 operator is independent of cavity 2, while cavity-2 operator expectations are influenced by cavity 1 (cf.~Appendix \ref{AppendixA}). 

In what follows, we show how this general recipe of balancing coherent and dissipative interactions can be used to construct an isolator and nonreciprocal quantum-limited amplifiers (both phase-preserving and phase-sensitive). The basic recipe here will in fact allow {\it any} factorizable cavity-cavity interaction to become directional, including nonlinear interactions (see Appendix \ref{AppendixA}).  It thus represents a powerful approach for constructing a wide variety of nonreciprocal behaviors.


\subsection{Unidirectional photon hopping: dissipative isolator}\label{subsec:IsolatorIntro}

We first discuss how our basic recipe can be used to obtain directional transmission between ports 1 and 2.  We want an
effective interaction between the two cavities which only allows photons to tunnel from cavity 1 to 2 (and not vice-versa).  This is precisely the kind of behavior described by standard cascaded quantum systems theory \cite{Carmichael1993,Gardiner1993,Gardiner2004}.  We show here how this fits into our general framework where directionality results from balancing coherent and dissipative interactions.  We also show how it can be simply realized using an auxiliary cavity or  reciprocal transmission line, and thus does not require an explicitly directional reservoir.

To obtain nonreciprocal tunneling between the cavities, we  first need to identify coherent and dissipative versions of a tunneling interaction.  The coherent version is simple:  choosing  
$\lambda = 0$ in Eq.~(\ref{Eq:Hint}), we obtain a standard hopping (or beam-splitter) Hamiltonian,
\begin{align}\label{Eq.:HoppHam}
	\hH_{\rm coh} \rightarrow &     J  \hat d_1^{\dag} \hat d_2 +  h.c.  \equiv \hH_{\rm hop} .
\end{align} 

For the dissipative version of a hopping interaction, we need a zero-temperature engineered reservoir that is able to absorb quanta
from either cavity; crucially there needs to be coherence between absorption of a photon from cavity 1 versus cavity 2.  The jump operator $\hat{z}$ in our
master equation Eq.~(\ref{Eq.:GeneralMaster})  thus needs to take the form
\begin{align}\label{Eq.:zhop}
	\hat{z} \rightarrow&    \hat d_1 + e^{i \varphi} \hat d_2 \equiv \hat{z}_{\rm hop}.
\end{align} 
The general master equation of Eq.~(\ref{Eq.:GeneralMaster}) thus reduces to
\begin{align}\label{Eq.:BSmasterequation}
 \frac{d}{dt} \hat \rho =& - i   \left[\hH_{\rm hop} , \hat \rho \right] 
			 + \Gamma \mathcal L [\hat d_1 + e^{i \varphi} \hat d_2 ] \hat{\rho}
			 +  \kappa   \sum_{j \in 1,2}   \mathcal L [\hat d_j ] \hat{\rho}.  
\end{align} 
The second term describes the dissipative hopping interaction:  the engineered reservoir can absorb a photon from either cavity 1 or cavity 2, and there is coherence between these possibilities (relative phase $\varphi$).  The rate for this process is $\Gamma$.  Note that via a gauge transformation, the phase $\varphi$ can be shifted into the phase of $J$.  We thus set $\varphi = 0$ in what follows, but keep $J$ complex.

\begin{figure} 
 \centering\includegraphics[width=0.45\textwidth]{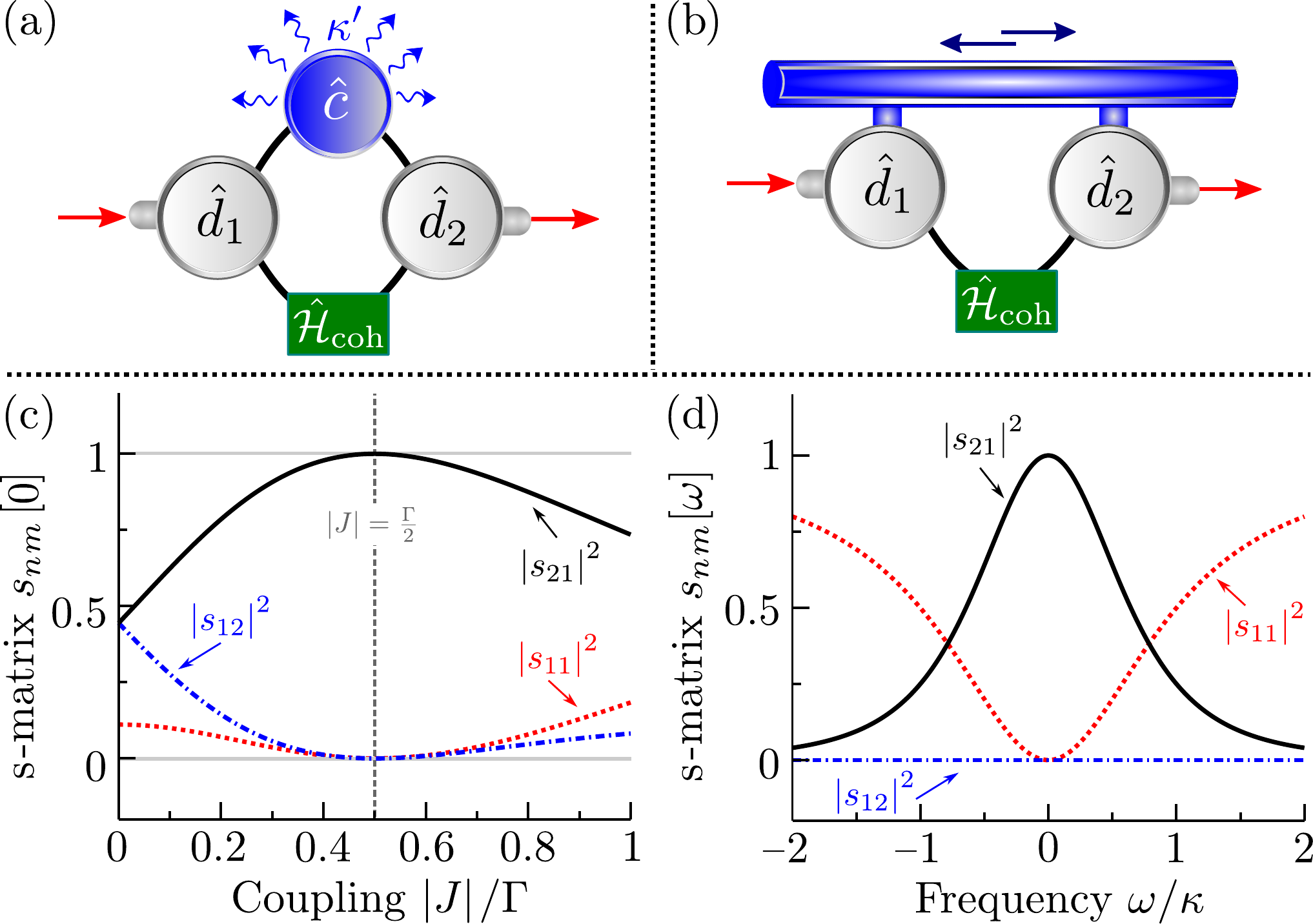}  
	\caption{
		(a) Realization of the engineered reservoir via an auxiliary cavity mode that is damped at a rate $\kappa^{\prime}$. 
		For strong damping $\kappa^{\prime} \gg \kappa$ this setup corresponds to a Markovian reservoir. 
		(b) Implementation based on a transmission line, which supports propagation of photons in both directions.  
		(c) Scattering matrix elements for the dissipative isolator setup at zero frequency, as a function of the
		coherent hopping $J$; the phase of $J$ is fixed so that $\arg(J/\Gamma) = \pi/2$.  When
		$J$ is tuned as per Eq.~(\ref{Eq:DirectionalHoppingCond}), the system only allows directional transmission 
		between cavities $1$ and $2$.  We have fixed the
		dissipative coupling strength $\Gamma$ to be equal to the cavity damping rate $\kappa$, and have taken the Markovian limit
		for the engineered reservoir ($\kappa' \gg \kappa$).   
		(d)  Scattering matrix elements as a function of frequency, when the directionality condition
		of  Eq.~(\ref{Eq:DirectionalHoppingCond}) is fulfilled.  In the Markovian limit, directionality holds over all frequencies.
		}
	\label{Fig:Isolator}
 \end{figure}

Before discussing how to engineer such a non-local dissipator, let us discuss the consequences.  Using Eq.~(\ref{Eq.:BSmasterequation}), the equations of motion for mode expectation values are
\begin{align}\label{Eq.:EoMexpIso}
	 \frac{d}{dt} \ev{\hat d_1} =& -  \frac{\kappa + \Gamma}{2}  \ev{\hat d_1} - \left[\frac{\Gamma}{2}   + i  J \right] \ev{\hat d_2},
	\nonumber \\ 
 	\frac{d}{dt} \ev{\hat d_2} =&  - \frac{\kappa + \Gamma}{2} \ev{\hat d_2} - \left[\frac{\Gamma}{2}    + i  J^{\ast}  \right] \ev{\hat d_1}.
\end{align}
Note that the engineered non-local dissipation in Eq.~(\ref{Eq.:BSmasterequation}) couples the two cavity lowering operators in an analogous manner to the coherent tunneling interaction $J$.  On a heuristic level, this is because the engineered reservoir gives rise to non-local damping:  the damping force on cavity $1$ depends on the amplitude of cavity $2$ (and vice-versa). If we only have the coherent hopping interaction (i.e.,~$\Gamma = 0$), or only have the dissipative interaction (i.e.,~$J=0$), the coupling between the cavities would be reciprocal. Note however that the coherent coupling involves $J$ in the first line of Eq.~(\ref{Eq.:EoMexpIso}), and $J^\ast$ in the second line.  The possibility thus emerges to have the two coupling terms {\it cancel} in one of the two equations.  
By setting, e.g., 
\begin{equation}
	J \overset{!}{=} i \frac{\Gamma}{2} , 
	\label{Eq:DirectionalHoppingCond}
\end{equation}	
 we obtain a unidirectional interaction: cavity 2 is driven by cavity 1, but not vice-versa (see Fig.~\ref{Fig:Isolator}(c)).

With this tuning of $J$, our master equation Eq.~(\ref{Eq.:BSmasterequation}) takes the standard form used in cascaded quantum systems theory \cite{Gardiner2004}:
\begin{align}\label{Eq.:BSmasterequationCascaded}
 \frac{d}{dt} \hat \rho  
	=  \left(\Gamma + \kappa \right) \sum_{n = 1,2} \mathcal L \left[ \hat d_n \right]  \hat{\rho}
   	-   \Gamma \left\{  \left[ \hat d_2^{\dag}, \hat d_1  \hat \rho\right]    - \left[\hat \rho \hat d_1^{\dag} ,\hat d_2\right] \right\}.   
\end{align}

We are most interested in the evolution of the extra-cavity fields, i.e.,~signals entering and leaving the two cavities via the coupling waveguides.  Treating our engineered dissipative reservoir using a Markovian oscillator bath (which is equivalent to the above Lindblad description), one can use standard input-output theory to calculate the relation between the input fields incident on the two cavities, 
$\hat d_{n, \rm in}$, and the output fields leaving the cavities, $\hat d_{n, \rm out}$ (see Sec.~\ref{Sec.:AuxMode} for details).  Using the input-output boundary condition  $\hat d_{n,\rm out} = \hat d_{n, \rm in} + \sqrt{\kappa} \hat d_n $ \cite{Gardiner1985,Clerk2010}, and letting
$\mathbf{D}[\omega]  =\left(  \hat d_{1 }[\omega] ,  \hat d_{2 }[\omega] \right)^T$, 
scattering between the cavity in/out fields is described by a $2 \times 2$ scattering matrix $\mathbf{s}[\omega]$:
\begin{align}
		\mathbf{D}_{\rm out}[\omega] = \mathbf{s}[\omega] \, \mathbf{D}_{\rm in}[\omega] +  \vec{\hat{\xi}}[\omega].
		\label{Eq:sDefn}
\end{align}
Here, $\hat{\xi}[\omega]$ describes (operator-valued) noise incident on the cavities from the engineered reservoir, and the zero frequency (i.e.,~on-resonance) scattering matrix is
\begin{align}\label{Eq.:SMatrixIso}
	\mathbf{s}[0] =  \left(
	\begin{array}{cc}
  		\frac{  \Gamma - \kappa }{\kappa + \Gamma}
		& 0 
	\\[2mm]
  		\frac{4 \kappa \Gamma}{(\kappa + \Gamma)^2}
		& \frac{  \Gamma - \kappa }{\kappa + \Gamma}
	\end{array}
	\right).
\end{align}

As expected, there is transmission from port $1$ to port $2$, but not vice versa.  
Note that $\mathbf{s}$ is  in general  not unitary, and hence the noise $\xi$ must be non-vanishing in order to preserve canonical commutators of the output fields; we discuss this noise in more detail in Sec.~\ref{Sec.:AuxMode}, showing that it can indeed have the minimal amount required by quantum mechanics.  We also show that the vanishing of $\mathbf{s}_{12}$ can be made to extend up to frequencies comparable to the relaxation rate of the engineered reservoir (i.e.,~much larger than $\kappa$), see Fig.~\ref{Fig:Isolator}(d).

Equation~(\ref{Eq.:SMatrixIso}) still does not have the ideal scattering matrix of an isolator \cite{Jalas2013}, as the incident signal on cavity $1$ could be partially reflected.  To suppress such reflections, we simply impedance match the system, i.e.,~tune $\Gamma = \kappa$.  We then obtain the ideal isolator scattering matrix (on resonance)
\begin{align}
\mathbf{s}[0] =  \left(
\begin{array}{cc}
  0 
& 0 
\\
  1 
& 0 
\end{array}
\right).
\end{align}
On a physical level, interference causes signals incident on cavity $2$ to be perfectly dumped into the dissipative reservoir.  Interference also ensures that signals incident on cavity $1$ never end up in this reservoir, but instead emerge from cavity $2$. 

We still have not specified {\it how} one obtains the required non-local dissipator.  The original works on cascaded quantum systems assumed an inherently nonreciprocal, unidirectional reservoir (i.e.,~a chiral transmission line), and 
then derived the effective master equation of Eq.~(\ref{Eq.:BSmasterequationCascaded}).  However, the above dynamics can be obtained {\it without} needing an explicitly directional reservoir.  One simple choice would be a one-dimensional transmission line,  cf.~Fig.~\ref{Fig:Isolator}(b), supporting both right-moving and left-moving modes, which couples to cavity $j$ at position $x_j$ 
\begin{equation}
	\label{Eq:SysBathWG}
	\hH_{\rm SB} = -\sqrt{\frac{\Gamma  \vwg}{2}} 
		\sum_{j=1,2} \left( \hat{d}_j^\dagger \left[ \hat{c}_R(x_j) + \hat{c}_L(x_j) \right] + h.c. \right),
\end{equation}
where $\hat{c}_{L,R}(x)$ denote the left and right moving fields in the waveguide, and $\vwg$ is the waveguide velocity.  For a suitable choice of $|x_1 - x_2|$, one again obtains the master equation of Eq.~(\ref{Eq.:GeneralMaster}) with jump operator $\hat{z}$ as per Eq.~(\ref{Eq.:zhop}).  Further details are provided in the Appendix \ref{AppendixB}.
	
Another simple implementation involves taking a damped auxiliary mode as the engineered reservoir (annihilation operator $\hat{c}$),  see Fig.~\ref{Fig:Isolator}(a), which interacts with the two principle modes via a Hamiltonian:
\begin{equation}
	\label{Eq:SysBath}
	\hH_{\rm SB} = J' \hat{c}^\dagger \left( \hat{d}_1 +  \hat{d}_2 \right) + h.c. .
\end{equation}
Such a quadratic interaction can be realized in a tunable fashion by starting with a three-wave mixing Hamiltonian and pumping one of the modes at an appropriate frequency; this is the same strategy used to implement the coherent direct interaction in Eq.~(\ref{Eq:Hint}) (see discussion following that equation).
As we show in Sec.~\ref{Sec.:AuxMode}, if the damping of the auxiliary mode $\kappa'$ is sufficiently large, it can be adiabatically eliminated, yielding the scattering matrix given above.  

For this particular realization, our isolator reduces to a three mode system with an asymmetric choice of damping rates.
Furthermore, the required phase of $J$ in the directionality matching condition of Eq.~(\ref{Eq:DirectionalHoppingCond}) corresponds to having the three mode system pierced by an effective magnetic flux of a quarter flux quantum.  
We stress that in many physical implementations, the couplings $J$ and $J'$ are tunable simply by controlling the amplitude and phases of the relevant pump modes. Thus, the directional interaction we finally obtain is not the result of having used an explicitly directional reservoir, but rather it
results from the control of relative phases in a driven system.  
Note that this three-mode realization of our dissipative isolator was also discussed
by Ranzani et al. \cite{Ranzani2014a}. It is also interesting that this three-mode implementation directly yields the scattering matrix of an ideal circulator (see Sec.~\ref{Sec.:AuxMode}); it is closely analogous to previous proposals for non-magnetic circulators \cite{Koch2010, Habraken2012, Ranzani2014a, Sliwa2015}.


\subsection{Directional phase-preserving quantum amplifier}\label{Sec.:NDPA}

We next use our general recipe to construct a nonreciprocal, phase-preserving amplifier, a topic that is of considerable interest to the superconducting qubit community \cite{Abdo2013b, Abdo2014,Ranzani2014a}.  We again consider a two-mode system as sketched in Fig.~\ref{fig:SketchCascaded}(b).  Our goal is a dynamics that leads to signals incident on cavity 1 emerging amplified from cavity 2, while at the same time, signals (and noise) incident on cavity 2 
are prevented from emerging from cavity 1.

Our basic recipe is the same as the previous subsection:  engineer both coherent and dissipative versions of the desired interaction, and then balance them to obtain directionality.  The coherent interaction needed corresponds to a non-degenerate parametric amplifier (NDPA), as obtained by setting $J=0$ in Eq.~(\ref{Eq:Hint}):
\begin{align}\label{Eq.:HamCohPA}
	 \hH_{\rm coh} \rightarrow&  \lambda  \hat d_1^{\dag} \hat d_2^{\dag} + \lambda^{\ast} \hat d_1  \hat d_2 \equiv \hH_{\rm PA}.
\end{align}
This textbook interaction results in the amplification of an input signal (or noise) incident on either cavity, in both transmission and reflection (see, e.g.,~Refs.~\onlinecite{Clerk2010, Walls2008}).

We next need to add the dissipative version of this NDPA interaction, as mediated by an appropriately chosen dissipative reservoir.  This kind of dissipative amplification was recently introduced in our previous work, Ref.~\cite{Metelmann2014}.  The dissipative reservoir now needs to be able to absorb photons from one cavity and to emit photons to the other, with coherence between these possibilities.  The jump operator $\hat{z}$ associated with the reservoir (cf.~Eq.~(\ref{Eq.:GeneralMaster})) thus needs to take the general form 
\begin{align}
	\label{Eq:zPA}
	\hat{z} \rightarrow&    \sqrt{2}   \left( \cos \theta \hat d_1 + e^{i \varphi} \sin \theta \hat d_2^\dagger \right) \equiv \hat{z}_{\rm PA},
\end{align} 
where the angle $\theta$ parametrizes the asymmetry between the two kinds of processes.
The relative phase $\varphi$ can again be gauged away into the phase of $\lambda$; we thus set it to zero in what follows.

With this choice of coherent Hamiltonian and dissipator, the two-cavity system is again described by the master equation Eq.~(\ref{Eq.:GeneralMaster}), with $\Gamma$ 
parametrizing the strength of the coupling to the engineered reservoir, and hence of the dissipative amplifier interaction.
To see clearly that the dissipation here 
leads to amplification, we consider the equations of motion for the means of lowering operators.  One obtains
\begin{align}
 		\frac{d}{dt} \ev{\hat d_1} =&   -  \frac{ \kappa + 2 \Gamma \cos^2 \theta }{2}  \ev{\hat d_1} 
						- \left[\frac{\Gamma}{2}   \sin 2 \theta    + i  \lambda  \right] \ev{\hat d_2^{\dag}},
	\nonumber \\ 
 		\frac{d}{dt} \ev{\hat d_2^{\dag}} =&  - \frac{\kappa - 2\Gamma \sin^2 \theta }{2} \ev{\hat d_2^{\dag}} 
						 + \left[\frac{\Gamma}{2}    \sin 2 \theta      + i  \lambda^{\ast} \right] \ev{\hat d_1}.
\end{align}
 \begin{figure} 
 \centering\includegraphics[width=0.48\textwidth]{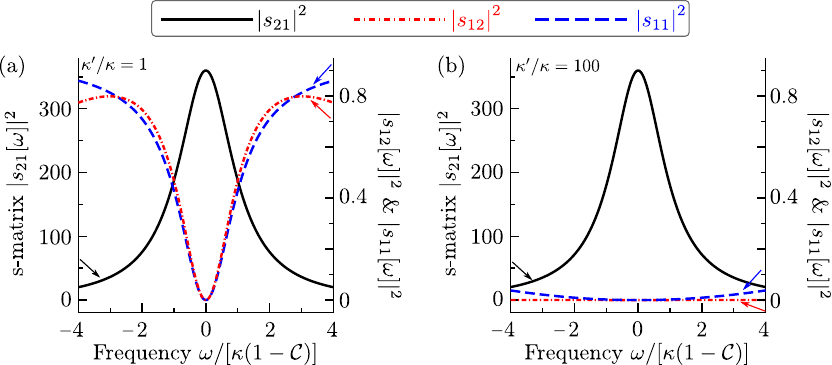}  
	\caption{
	s-matrix elements of the directional, phase preserving amplifier, as a function of scaled frequency; 
	the cooperativity $\mathcal{C} \equiv \Gamma / \kappa = 0.95$, where $\Gamma$ is the dissipative interaction strength, and
	$\kappa$ is the damping rate of cavities $1$ and $2$.
	(a) Auxiliary mode damping $\kappa' = \kappa$, indicating a strong deviation from  the Markovian limit;
	while perfect isolation exists at $\omega = 0$, it is rapidly lost for non-zero frequencies.
	(b) Auxiliary mode damping $\kappa' = 100 \kappa$, closer to the Markovian limit.  The directionality is much better at finite frequencies, while the
	gain is unchanged.}
	\label{Fig:NDPA}
 \end{figure}
The crucial terms behind the amplification are the last term in each line, which cause $\hat{d}_1$ to be driven by $\hat{d}_2^\dagger$ and vice-versa.  Again, both the coherent interaction and the dissipative interaction give rise to such terms; each interaction thus facilitates amplification that can be quantum limited, but that is not directional \cite{Metelmann2014}.  To obtain a unidirectional interaction, we again simply tune the amplitude and phase of the coherent interaction with respect to the dissipative interaction, so as to cancel the coupling term in the first equation, i.e.,
\begin{equation}
	\lambda \overset{!}{=}  i  \frac{\Gamma}{2}   \sin 2 \theta     .
	\label{Eq:DirectionalNDPACond}
\end{equation}

To see that this choice gives the desired behavior of the output fields, we model the dissipative bath as a Markov reservoir, and calculate the scattering matrix for the system using input-output theory.
Letting  $\mathbf{D}_{\rm in/out}[\omega]  =\left(  \hat d_{1,\rm in/out }^{\PD}[\omega] ,  \hat d_{2,\rm in/out }^{\dag}[\omega] \right)^T$, the input-output relations take the form of Eq.~(\ref{Eq:sDefn}).  $\vec{\hat{\xi}}$ again describes noise incident from the engineered reservoir, while the $2 \times 2$ scattering matrix $\mathbf{s}$ takes the following explicitly nonreciprocal form at zero frequency
 \begin{align}
 	\label{Eq:NDPASMatrix}
	\mathbf{s}[0] =  \left(
	\begin{array}{cc}
  		\frac{  2 \mathcal C  \cos^2 \theta  -  1  }{   2 \mathcal C  \cos^2 \theta + 1 }
			& 0 
		\\[2mm]
  			\frac{4  \mathcal C   \sin 2 \theta  }
				     { \left[ 2  \mathcal C  \cos^2 \theta + 1 \right] \left[ 2 \mathcal C  \sin^2 \theta - 1\right] } 
			&  \frac{ 2 \mathcal C  \sin^2 \theta  + 1}{2 \mathcal C \sin^2 \theta - 1}
	\end{array}
	\right).
\end{align}
Here, the cooperativity is given as $\mathcal C =  \frac{\Gamma}{\kappa} $.

If we further tune $\theta$ so that 
\begin{equation}
	\cos^2 \theta  \overset{!}{=}  1/(2  \mathcal C ),
	\label{Eq:NDPAImpedanceMatchCond}
\end{equation}
(possible as long as $ \mathcal C  > 1/2$), we cancel all reflections of input signals incident on cavity $1$. With this tuning, the scattering matrix becomes
 \begin{align}
 	\label{Eq:NDPASMatrixIM}
	\mathbf{s}[0] =  \left(
	\begin{array}{cc}
			0 & 0 
		\\ 
  			\sqrt{ \mathcal G} &  \sqrt{ \mathcal G + 1}
	\end{array}
	\right),
\end{align}
with $ \mathcal G  = \frac{  2 \mathcal C - 1   }{(\mathcal C  - 1)^2     }$.  
As claimed, we have a scattering matrix describing nonreciprocal, phase-preserving amplification, with a gain that diverges as $\mathcal{C}$ approaches 1.
Signals incident on cavity $1$ are never reflected, and emerge from cavity $2$ with an amplitude gain $s_{21} = \sqrt{\mathcal G}$, 
whereas signals incident on cavity $2$ do not emerge at the output from cavity $1$.
The system exhibits a standard parametric instability when $\mathcal{C} > 1$ (analogous to the instability in a standard, coherent NDPA).

The frequency dependence of the scattering coefficients is discussed in Sec.~\ref{Sec.:NDPAProp}. 
Strikingly, the directionality property $s_{12}[\omega] = 0$ holds for all frequencies for which the Markovian bath approximation is valid.  The system is limited by a standard gain-bandwidth constraint (in contrast to the purely dissipative amplification process, which is has no such constraint \cite{Metelmann2014}).  
We also discuss the added noise of the amplifier in Sec.~\ref{Sec.:NDPAProp}, showing that it is quantum limited in the large gain limit as long as there is no thermal noise incident on cavity $2$;
surprisingly, the engineered reservoir need not be at zero-temperature.

While there are many ways to realize the engineered reservoir used in this scheme, the simplest choice is a damped third auxiliary mode, see
Sec.~\ref{Sec.:NDPAProp}.  With this particular choice, our scheme reduces to the 3-cavity amplifier discussed by Ranzani and Aumentado in Ref.~\onlinecite{Ranzani2014a}.  Our analysis thus generalizes this scheme, and provides insight into the underlying mechanism.
It also shows the crucial importance of having the auxiliary mode damping $\kappa'$ be much larger than that of the principle modes; in this 
Markovian limit, one has directionality over the full amplification bandwidth (see Fig.~\ref{Fig:NDPA}).


\subsection{Directional phase-sensitive amplifier}\label{Sec.:DPA}

As a third application of our recipe for nonreciprocity, we construct a phase-sensitive amplifier.  Phase-sensitive amplifiers only measure a single quadrature of an incident signal; as a result, quantum mechanics allows them to amplify without adding any added noise \cite{Caves1982,Clerk2010}.  Our general approach allows one to construct a nonreciprocal and noiseless version of such an amplifier, again using the two-cavity-plus-reservoir setup in Fig.~\ref{fig:SketchCascaded}(b).  The resulting amplifier has another striking advantage over a standard paramp:  it does not suffer from any fundamental gain-bandwidth limitation, a point we discuss more fully in Sec.~\ref{Sec.:DPAMode}.

As before, the first step is to construct a coherent interaction that gives the desired amplification.  The standard choice would be a degenerate parametric amplifier (DPA) Hamiltonian involving just a single mode, of the form $\hH_{\rm int} = \lambda \hat{d} \hat{d} + \lambda^* \hat{d}^\dagger \hat{d}^\dagger$.  In contrast, to be able to implement our recipe for directionality, we want an interaction that couples {\it two} modes.   

 \begin{figure}[t]
 \centering\includegraphics[width=0.5\textwidth]{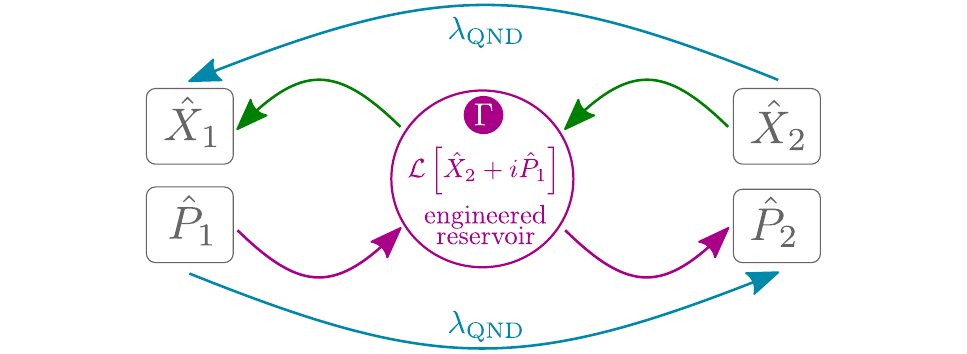} 
	\caption{Schematic illustrating the directional phase-sensitive amplifier.  The coherent QND Hamiltonian of Eq.~(\ref{Eq:HamCohAmp}) causes the
	cavity-1 $P$ quadrature to drive the cavity-2 $P$ quadrature, and the cavity-2 $X$ quadrature to drive the cavity-1 $X$ quadrature (blue arrows); 
	there is gain associated with each of these drivings, cf.~Eqs.~(\ref{Eq.:EoMexpectDPA}).  
	The engineered reservoir (jump operator described by Eq.~(\ref{Eq:zQND})) also mediates the same drivings (green and magenta arrows), 
	again with gain.  By balancing these interactions, one can cancel the $X_2 \rightarrow X_1$ driving, resulting in directional amplification.
	 }
	 \label{Fig:DPASchematic}
 \end{figure}

Surprisingly, there is a simple coherent two-mode interaction which does the job and which yields ideal amplification properties (zero added noise, no gain-bandwidth limitation).  One needs to use the kind of quantum non-demolition (QND) interaction discussed extensively in the context of Gaussian cluster-state generation \cite{Zhang2006,Menicucci2006,Weedbrook2012}.  Suppose we want to amplify the $P$ quadrature of cavity 1, i.e.,~the operator $\hat{P}_1 = 
-i(\hat{d}_1^{\PD} - \hat{d}_1^\dagger) / \sqrt{2}$.  We then use an interaction Hamiltonian that commutes with this operator, but that takes information in the $P_1$ quadrature and dumps it into a cavity $2$ quadrature (i.e.,~$P_2$).  
The required coherent Hamiltonian is obtained by setting 
$J = \lambda = i \lambda_{\rm QND}/2 \ (\lambda_{\rm QND} \in \mathbb{R})$ in Eq.~(\ref{Eq:Hint}), i.e.,~
\begin{align}\label{Eq:HamCohAmp}
	 \hH_{\rm coh} \rightarrow& \lambda_{\rm QND} \ \hat P_1 \hat X_2 \equiv \hH_{\rm QND},
\end{align}
with $\hat X_2 = \left( \hat d_2^{\PD} + \hat d_2^{\dag} \right)/\sqrt{2}$.  
It is straightforward to see from the Heisenberg equations of motion that $\hH_{\rm QND}$ causes $P_2$ to be driven by $P_1$, and hence $P_2$ will contain information on $P_1$ (see Fig.~\ref{Fig:DPASchematic}).  The same holds for the extra-cavity fields:  the $P$ quadrature of a signal incident on cavity $1$ will emerge in the $P$ quadrature of a signal leaving cavity $2$.

Note that $\hat P_1$ and $\hat X_2$ are QND variables:  they commute with the Hamiltonian in Eq.~(\ref{Eq:HamCohAmp}), and are undisturbed by the amplification process.  It follows that there is no possibility of feedback in this system, and hence the system is stable irrespective of the value of $\lambda_{\rm QND}$.  By increasing $\lambda_{\rm QND}$, one can thus achieve increasing amounts of phase-sensitive gain.  Furthermore, as the amplification mechanism here does not involve coming close to an instability, the amplification bandwidth is always $\sim \kappa$, irrespective of the gain.

Following our recipe for directionality, we next need to construct the dissipative counterpart to the coherent interaction in Eq.~(\ref{Eq:HamCohAmp}).  We need the jump operator $\hat{z}$ characterizing
the engineered reservoir to also preserve the QND structure of the coherent Hamiltonian.  Taking $u_1 =  u_2 = v_2 = \sqrt{2}$ and $v_1 = -\sqrt{2}$ in Eq.~(\ref{Eq:zDefn}) yields
\begin{align}
	\label{Eq:zQND}
	\hat{z} \rightarrow&    \hat{X}_2 + i \hat{P}_1  \equiv \hat{z}_{\rm QND}.
\end{align} 
This dissipative interaction is the counterpart of the coherent interaction in Eq.~(\ref{Eq:HamCohAmp}):  with this choice of $\hat{z}$, the dissipative terms in Eq.~(\ref{Eq.:GeneralMaster}) alone
lead to amplification of the $P$ quadrature of signals incident on cavity 1. 
The heuristic interpretation of this dissipative amplification is similar to that presented in Ref.~\onlinecite{Metelmann2014} for the phase-preserving case:  the engineered reservoir ``measures" the QND quadrature $\hat{P}_1$, and then dumps this information into the non-QND quadrature $\hat{P}_2$
(see Fig.~\ref{Fig:DPASchematic}, as well as Sec.~\ref{Sec.:DPAMode}).

With these choices for $\hH_{\rm coh}$ and $\hat{z}$ in Eq.~(\ref{Eq.:GeneralMaster}), we have both coherent and dissipative phase-sensitive amplifying interactions between the cavities.  Using this master equation, the equations of motions for the quadrature means have the expected form:
\begin{align}\label{Eq.:EoMexpectDPA}
 \frac{d}{dt} \ev{\hat P_1} =&  - \frac{\kappa}{2} \ev{\hat P_1}   ,
\nonumber \\  
 \frac{d}{dt} \ev{\hat X_2} =&  - \frac{\kappa}{2} \ev{\hat X_2}    ,
   \nonumber \\ 
 \frac{d}{dt}\ev{ \hat X_1} =& - \frac{\kappa}{2} \ev{\hat X_1}   + \left[ \lambda_{\rm QND}   - \Gamma  \right] \ev{\hat X_2} ,
\nonumber \\ 
 \frac{d}{dt} \ev{\hat P_2} =& - \frac{\kappa}{2} \ev{\hat P_2}  -  \left[ \lambda_{\rm QND}  +  \Gamma  \right] \ev{\hat P_1}. 
\end{align}
$P_1$ and $X_2$ are QND variables and thus undisturbed by either interaction.  In contrast, both interactions cause $P_2$ to become an amplified copy of $P_1$.

We can now finally apply the last step of our general recipe:  balance the dissipative and coherent interactions to break reciprocity.  This simply involves setting
\begin{equation}
	\Gamma  \overset{!}{=} \lambda_{\rm QND},
	\label{Eq:DPADirectionalCond} 
\end{equation}
which ensures that cavity $1$ is insensitive to the state of cavity $2$.

 \begin{figure}
 \centering\includegraphics[width=0.5\textwidth]{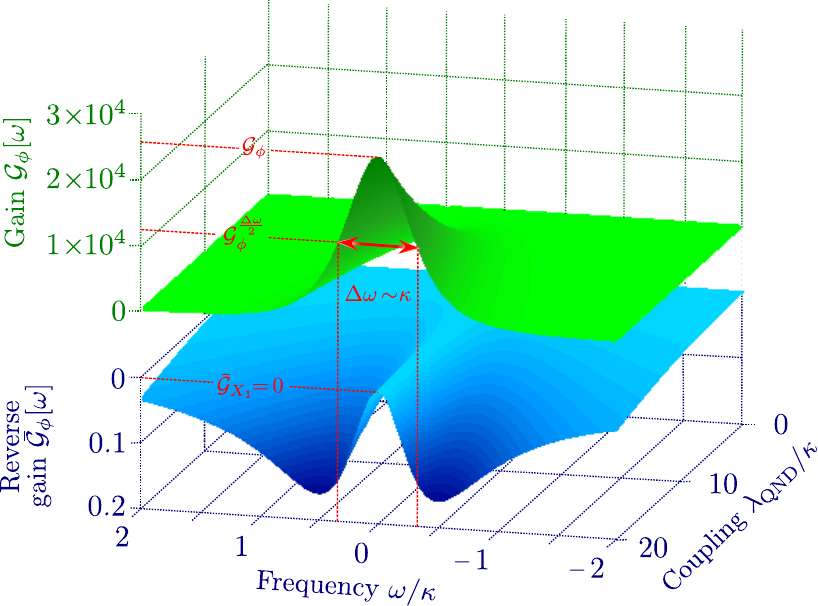} 
	\caption{Gain $\mathcal{G}_{\phi}[\omega]$ and reverse gain $\bar{\mathcal{G}}_{\phi}[\omega]$ of the directional phase-sensitive amplifier, 
	plotted as a function of signal frequency $\omega$ and coherent coupling strength $\lambda_{\rm QND}$ (cf.~Eq.~(\ref{Eq:HamCohAmp})), assuming
	that the dissipative coupling $\Gamma$ always satisfies the matching condition $\Gamma = \lambda_{\rm QND}$
	(cf.~Eq.~(\ref{Eq:DPADirectionalCond})).
	$\mathcal{G}_{\phi}[\omega]$ describes the amplification in transmission of signals incident on the cavity-1 $P$ quadrature, while
	$\bar{\mathcal{G}}_{\phi}[\omega]$ describes the amplification in transmission of signals incident on the cavity-2 $X$ quadrature.	  
	We have taken the engineered reservoir to be an auxiliary cavity mode with damping rate 
	$\kappa^{\prime}/\kappa = 100$ (see Sec.~\ref{Sec.:DPAMode}).  In this limit, deviations 
	from the Markovian-reservoir approximation are small.}
	\label{Fig:DPA3D}
 \end{figure}

Finally, we are as usual interested in the behavior of the output fields from the cavity.  Treating the engineered reservoir as a Markovian bath and using input-output theory, we can again calculate the scattering matrix of the system.  Writing this matrix in a quadrature representation,  we find that on-resonance (i.e.,~at zero-frequency) $\mathbf{Z}_{\rm out}  = \mathbf{s}  \ \mathbf{Z}_{\rm in} + \vec{\hat{\xi}}$ with
\begin{align}\label{Eq.DirAmpOutputShort}
\mathbf{s}[0] =  \left(
\begin{array}{cccc}
  -1 
& 0 
& 0
& 0
\\
  0 
& -1 
& 0
& 0
\\
  0 
& 0
& -1
& 0
\\
  0 
&  \sqrt{\mathcal G_\phi} 
& 0
& -1
\end{array}
\right),
\hspace{0.3cm}
\mathbf{Z}  =
\left(
\begin{array}{c}
 \hat X_{1}
\\
  \hat P_{1}
\\
  \hat X_{2}
\\
  \hat P_{2}
\end{array}
\right).
\end{align}
Here, the zero-frequency amplitude gain is given by $\sqrt{\mathcal G_\phi} =  \frac{8\lambda_{\rm QND}}{\kappa}$. 

The input-output relations in Eqs.~(\ref{Eq.DirAmpOutputShort}) describe an ideal directional degenerate amplifier:
the $P$ quadrature of signals incident on cavity $1$ emerge with gain from cavity $2$, whereas signals or noise incident on cavity $2$ never emerge from cavity $1$.  
Note further that there is no unwanted amplification in reflection of incident signals and noise.
The amplifier also has several other remarkable properties:  it is quantum limited (i.e., no added noise in the large gain limit), and does not suffer from any fundamental gain-bandwidth limitation. 
The directionality is also maintained over a large range of frequencies (see Fig.~\ref{Fig:DPA3D}).
These properties (along with the possibility of eliminating unwanted reflections) are discussed in more detail in Sec.~\ref{Sec.:DPAMode}.


\section{Noise, bandwidth and three-cavity implementation}\label{Sec.:Three:Implementations}


\subsection{Dissipative isolator: additional details}\label{Sec.:AuxMode}


\subsubsection{Auxiliary-cavity implementation of the engineered reservoir}\label{subsubsec:AuxMode}

To demystify the engineered reservoirs used in our schemes, we provide more details here on the simplest possible realization: 
a damped auxiliary cavity mode. In the limit where the damping rate $\kappa'$ of this auxiliary mode 
is large, this model describes a general Markovian reservoir.  We stress that this setup is just one of many ways to implement the necessary dissipative dynamics.
In Appendix \ref{AppendixB}, we explicitly show how coupling two cavities to a (non-directional) one-dimensional transmission line or waveguide also generates the needed dissipative dynamics.

Consider the dissipative isolator described by the master equation in  Eq.~(\ref{Eq.:BSmasterequation}), and take the engineered reservoir to be an auxiliary mode with lowering operator $\hat{c}$ which is damped at rate $\kappa'$ by a coupling to a Markovian reservoir.  $1/\kappa'$ will act as the correlation time of our engineered reservoir. As discussed in Sec.~\ref{subsec:IsolatorIntro}, we need this auxiliary mode (i.e.,~the engineered reservoir) to interact with the principle modes via the interaction Hamiltonian in Eq.~(\ref{Eq:SysBath}).  The simplest limit is where $\kappa'$ is much larger than all other frequency scales; in this limit, the $\hat{c}$ mode will itself act as a Markovian reservoir for the system modes $\hat{d}_1, \hat{d}_2$.  One could then recover the master equation of Eq.~(\ref{Eq.:BSmasterequation}) using standard adiabatic elimination techniques \cite{Gardiner2004}.  

Alternatively, one can eliminate the auxiliary mode within a Heisenberg-Langevin formalism, using the coherent Hamiltonian $\hH = \hH_{\rm hop} + \hH_{\rm SB}$.  
Solving the equation of motion for $\hat{c}$ in the large-damping (adiabatic) limit yields
\begin{align}
  \hat c   =&  - \frac{2}{\sqrt{\kappa^{\prime}}} \hat c_{ \rm in}  - i \frac{2 J^{\prime}}{\kappa^{\prime}}  \left(\hat d_1 +  \hat d_2 \right) ,
\end{align}
where all operators are evaluated at the same time, and $\hat c_{ \rm in}$ describes thermal and vacuum fluctuations stemming from the mode's internal dissipation.  Substituting this equation into the equations of motion for the principle cavity operators $\hat{d}_j$ then yields:
\begin{align}
 \frac{d}{dt} \hat d_1 =& - \frac{\kappa + \Gamma}{2}     \hat d_1
			- \sqrt{\kappa} \hat d_{1,\rm in}
			+ i \sqrt{\Gamma} \hat c_{ \rm in}  
			-  \left[  \frac{\Gamma}{2}  + i J \right] \hat d_2  ,
\nonumber \\   
 \frac{d}{dt} \hat d_2 =& - \frac{\kappa + \Gamma}{2} \hat d_2 
			- \sqrt{\kappa} \hat d_{2,\rm in}   
			+ i\sqrt{\Gamma}  \hat c_{ \rm in}  
			-  \left[ \frac{\Gamma}{2}  +  i J^{\ast}\right] \hat d_1   ,
	\label{Eqs:AuxCavEOM}
\end{align}
where we take $J^{\prime} \in \mathbb{R}$ without loss of generality, and define $\Gamma \equiv  4  J^{\prime 2}  / \kappa^{\prime} $.
Taking average values, we recover the master-equation result of  Eq.~(\ref{Eq.:EoMexpIso}). 

Using the Heisenberg-Langevin approach, we can now calculate the {\it full} scattering matrix for the system which includes the scattering of noise incident from the 
engineered reservoir. Letting $\mathbf{Y}[\omega] = \left(\hat d_{1}[\omega],\hat d_{2}[\omega],\hat c[\omega]\right)^T$, the full scattering relations take the form
$\mathbf{Y}_{\rm out}[\omega] = \tilde{\mathbf{s}}[\omega] \ \mathbf{Y}_{\rm in}[\omega]$. 
Consider first the Markovian limit, where $\kappa' \gg \omega, \kappa, \Gamma$.  Assuming that the system has been tuned to satisfy 
both the directionality condition $J = i  \frac{\Gamma}{2}$ (cf.~Eq.~(\ref{Eq:DirectionalHoppingCond})) and the impedance matching condition $\kappa = \Gamma$, the full scattering matrix in this
limit is
\begin{equation}   
\tilde{\mathbf{s}}[\omega] =
\left(
\mkern-5mu
\begin{tikzpicture}[baseline=-.65ex]
\matrix[
  matrix of math nodes,
  column sep=0.5ex,
] (m)
{
   \frac{ - i \frac{ \omega}{\kappa} }
	{1 - i \frac{\omega}{\kappa} } 
&   
	0 
&  \frac{ i }
	{1    - i \frac{  \omega}{\kappa} } 
\\
   \frac{1  }
	{ \left(1  - i \frac{ \omega}{\kappa} \right)^2}  
& 
   \frac{ - i \frac{ \omega}{\kappa} }
	{1   - i \frac{ \omega}{\kappa}}
&
   \frac{ \frac{ \omega}{\kappa} }
	{\left(1  - i \frac{ \omega}{\kappa}\right)^2}
\\
   \frac{  \frac{ \omega}{\kappa}}
	{\left(1   - i \frac{ \omega}{\kappa} \right)^2} 
& 
   \frac{  i    }
	{1    - i \frac{  \omega}{\kappa}} 
&
   \frac{\left(     \frac{  \omega^2}{\kappa^2} \right) }
	{\left(  1 - i \frac{  \omega}{\kappa} \right)^2} 
\\
};
\draw[dashed]
  ([xshift=2.5ex]m-1-2.north east) -- ([xshift=0.5ex]m-2-2.south east);
\draw[dashed]
  ( m-2-1.south west) -- ([yshift=-0.65ex]m-2-2.south east); 
\end{tikzpicture}
\mkern-5mu
\right) 
+ \mathcal O\left[\frac{1}{\kappa^{\prime}} \right].
\end{equation}
The upper left $2\times 2$ matrix is the scattering matrix $\mathbf{s}$ for the reduced, two-mode system, cf.~Eq.(\ref{Eq.:SMatrixIso}).  The elements $\tilde{s}_{13}$ and $\tilde{s}_{23}$ describe the scattering of noise from the engineered reservoir to the main cavity modes.  This then explicitly yields the noise operator in Eq.~(\ref{Eq:sDefn}) as $\vec{\hat{\xi}} = [ \tilde{s}_{13}, \tilde{s}_{23} ]^T \hat{c}_{\rm in}$.  We see that directionality holds for all frequencies in this Markovian limit, i.e., $\tilde{s}_{12}[\omega] = 0$.  In contrast, the impedance matching (which ensures no reflections at the input of cavity $1$) only holds for $\omega \ll \kappa$.

Finally, note that at zero frequency, the full scattering matrix becomes:
\begin{equation}  \label{Eq.:MatrixIsoAuxMode}
\tilde{\mathbf{s}}[0] =
\left(
\mkern-5mu
\begin{tikzpicture}[baseline=-.65ex]
\matrix[
  matrix of math nodes,
  column sep=1ex,
] (m)
{
0 & 0 & i\\
1 & 0 & 0\\
0 & i & 0\\
};
\draw[dashed]
  ([xshift=0.5ex]m-1-2.north east) -- ([xshift=0.5ex]m-2-2.south east);
\draw[dashed]
  (m-2-1.south west) -- (m-2-2.south east); 
\end{tikzpicture}
\mkern-5mu
\right) .
\end{equation}
In this ideal case, signals incident on cavity $2$ are perfectly transmitted to the engineered reservoir, while the input field on the reservoir (i.e.,~the $\hat{c}$ mode) is perfectly transmitted to mode $1$. If the engineered reservoir is at zero temperature, we see that the output from cavity 1 is simply vacuum noise.  Amusingly, the above unitary scattering matrix is that of a perfect circulator:  the effective magnetic field associated with the phase of $J$ breaks the degeneracy of right and left circulating eigenmodes of the coherent three-mode hopping Hamiltonian. 
In the case of symmetric decay rates, i.e., $\kappa^{\prime} = \kappa$,  this kind of circulator
has been discussed in the context of superconducting circuit setups \cite{Koch2010, Ranzani2014a} and just recently experimentally demonstrated by Sliwa and co-workers \cite{Sliwa2015}. An analogous circulator for phonons has been discussed in the context of optomechanics \cite{Habraken2012}.  


\subsubsection{Non-Markovian corrections}\label{Subsec:IsolatorNonMarkov}

We can also consider deviations from the Markovian limit, where 
the internal damping rate of the engineered reservoir $\kappa'$ is not arbitrarily large.  The scattering matrix follows simply from solving the full (linear) Langevin equations without any adiabatic assumption.  We quote only the results for the forward and reverse transmission probabilities, again assuming that 
the directionality and impedance matching conditions have been met. We find 

\begin{align}
|\tilde{s}_{21}[\omega]|^2 =& \frac{       \left(1 +   \frac{ \omega^2}{\kappa^{\prime 2}}    \right)}
			    {    \left[\frac{\omega^2}{\kappa^{\prime 2}} \left(1+ \frac{4 \omega^4}{\kappa^4}\right) -\frac{4\omega^4}{\kappa^3\kappa^{\prime  }}
					+\left(1+\frac{\omega ^2}{\kappa^2}\right)^2  \right]},
\nonumber \\
 |\tilde{s}_{12}[\omega]|^2 =& \frac{      \frac{ \omega^2}{\kappa^{\prime 2}}  }
			{    \left[\frac{\omega^2}{\kappa^{\prime 2}} \left(1+ \frac{4 \omega^4}{\kappa^4}\right)  -    \frac{4\omega^4}{\kappa^3\kappa^{\prime  }}
				   +\left(1+\frac{\omega ^2}{\kappa^2}\right)^2  \right]}.
\end{align}
One clearly sees that the directionality only holds for frequencies that are small compared to the inverse correlation time $1/\kappa'$ of the 
reservoir:  for small $\omega$, $|\tilde{s}_{12}[\omega] |^2 \propto \omega^2 / \kappa'^2$.
For non-zero $\omega / \kappa'$, the engineered reservoir gives rise to both dissipative and coherent interactions.  
The extra induced coherent interaction ruins the directionality matching condition of Eq.~(\ref{Eq:DirectionalHoppingCond}), leading to a lack of perfect isolation.


\subsection{Directional phase-preserving amplifier: additional details}\label{Sec.:NDPAProp}


\subsubsection{Bandwidth and non-Markovian effects}

We return now to the setup presented for directional amplification in Sec.~\ref{Sec.:NDPA}.  As in the previous section, 
we will investigate the frequency-dependent behavior of the system
using an auxiliary damped cavity mode $\hat{c}$ to represent the engineered reservoir.  With this choice, the system is analogous to that studied
in Ref.~\onlinecite{Ranzani2014a} , which was recently implemented in a superconducting circuit experiment \cite{Sliwa2015}.  We emphasize the importance of having a large damping rate $\kappa'$ of the auxiliary mode, thus complementing the discussion in Ref.~\onlinecite{Ranzani2014a}.

For phase-preserving amplification, the coherent interaction between the principle modes $\hat{d}_1,\hat{d}_2$ has the NDPA form of Eq.~(\ref{Eq.:HamCohPA}).  To obtain the correct dissipative interaction, the coupling $\hH_{\rm SB}$ between the principle cavity modes and the
auxiliary mode should have the form
\begin{equation}
	\label{Eq:SysBathPA}
	\hH_{\rm SB} = \sqrt{2} \lambda' \hat{c}^\dagger \left(  \cos \theta \, \hat{d}_1 + \sin \theta \, \hat{d}_2^\dagger \right) + h.c. .
\end{equation}
Taking the large $\kappa'$ limit and using standard adiabatic elimination techniques, one recovers the master equation described by Eqs.~(\ref{Eq.:GeneralMaster}) and (\ref{Eq:zPA}), with $\Gamma = 4 \lambda'^2 / \kappa'$.

One can again solve the full Heisenberg-Langevin equations to obtain the full $3 \times 3$ scattering matrix for the system at all frequencies.
If we tune the couplings to satisfy the directionality condition of Eq.~(\ref{Eq:DirectionalNDPACond}) and the impedance matching condition
of Eq.~(\ref{Eq:NDPAImpedanceMatchCond}),
 the ``forward photon number gain" of the amplifier takes the form  
 \begin{align}
 	\mathcal G[\omega] \equiv  \left| s_{21}[\omega] \right|^2  =&  
	\frac{ \left(2 \mathcal C -1\right)      }
		       {\left[\frac{ \omega^2}{\kappa^2}+1\right] \left[\left(  \mathcal C -1\right)^2 + \frac{ \omega^2}{\kappa^2}\right]}  
		+\mathcal O \left[ \frac{\omega}{\kappa^{\prime}}\right].
\end{align}
The corresponding reverse photon number gain (which we ideally want to vanish) is given by
\begin{align}
 	\bar{\mathcal G}[\omega] \equiv  \left| s_{12}[\omega] \right|^2 =    
	 \mathcal G[\omega]  \frac{\omega^2}{\kappa^{\prime 2}} +\mathcal O \left[ \frac{\omega^3}{\kappa^{\prime 3}}\right]   .
	\label{Eq:NDPAReverseGain}
\end{align}

Consider first the limit where the engineered reservoir is effectively Markovian, $\omega / \kappa' \rightarrow 0$.  
The reverse gain always vanishes, while the zero frequency forward gain
$\mathcal G[0]$ is controlled by $\mathcal C$, and diverges as $\mathcal C \rightarrow 1$; the system is unstable for larger $\mathcal C$.   In the large gain limit, $\mathcal G[\omega]$ is a Lorentzian as a function of frequency, with a bandwidth
$\Delta \omega = 2\kappa (1 - \mathcal C)$ that decreases as one increases the gain..  
The amplifier has a finite gain-bandwidth limitation just like a standard cavity-based NDPA 
(i.e., the product $\sqrt{\mathcal G[0]} \Delta \omega$ is fixed) \cite{Ranzani2014a}.   
Note that the dissipative parametric interaction on its own suffers from no such limitation \cite{Metelmann2014}, but is of course not directional.  Directionality is thus obtained by introducing a coherent NDPA interaction, with the price that this interaction naturally leads to a conventional gain-bandwidth limit.

Turning to the non-Markovian effects, we see from Eq.~(\ref{Eq:NDPAReverseGain}) 
that for finite $\omega / \kappa'$, the reverse gain is non-zero, implying that directionality is lost; this is also depicted in Fig.~\ref{Fig:NDPA}.  
The loss of directionality here is analogous to what happens in the directional isolator, and occurs for the same basic physical reason:  for finite $\omega / \kappa'$, the engineered reservoir also induces a coherent interaction between the two modes, and hence the perfect matching of coherent and dissipative interactions needed for directionality is lost.


\subsubsection{Added noise and quantum-limited behavior}

In addition to directionality, for many applications it is crucial that our amplifier reaches the fundamental quantum limit on its added noise.  
This limit corresponds to adding noise equivalent to half a quanta at the input, $\bar n_{  \rm add} \geq 1/2$ \cite{Caves1982}.  The added noise follows directly
from the full scattering matrix $\tilde{s}$, and will have contributions both from noise incident on cavity 2 that is reflected, and noise emerging from the engineered reservoir (i.e.,~the auxiliary $\hat{c}$ mode).  Assuming that the impedance matching and directionality conditions have been fulfilled, 
and letting $\bar{n}_{d_2}^T$ and $\bar{n}_c^T$ represent the thermal occupancies (respectively) of these two noise sources, we find: 
\begin{align}
	 \bar n_{\rm add}[0] =   \left(\frac{1}{2} + \bar n_{d_2}^T \right) \left[1 +\frac{1}{\mathcal G [0]} \right].
\end{align}
Thus, in the large gain limit, the added noise is quantum limited as long as there is no thermal noise incident upon cavity 2 (i.e., $n_{d_2}^T=0$)
\cite{Ranzani2014a}. Remarkably, thermal noise in the engineered reservoir does not prevent one from reaching the quantum limit; similar behavior is found in a purely dissipative (non-directional) phase-preserving amplifier \cite{Metelmann2014}.

While not relevant to the quantum limit, from a practical standpoint one also wants the noise leaving cavity $1$ to be small (so as not to damage the signal source).  Using our scattering matrix, it is straightforward to calculate the noise of the output field from cavity one.  Characterizing this noise by an effective thermal occupancy $\bar{n}^T_{1,\rm out}$, we find at zero frequency:
\begin{align}
	\bar{n}^T_{1,\rm out} = \bar n_c^T.
\end{align}
Thus, while thermal noise in the engineered reservoir does spoil quantum limited performance, this noise does show up in the output of cavity $1$.


\subsection{Directional phase sensitive amplifier: additional details}\label{Sec.:DPAMode}


\subsubsection{Full scattering matrix and impedance matching}

We now turn attention to our scheme of Sec.~\ref{Sec.:DPA} for directional and noiseless single-quadrature amplification.  As discussed in that section, 
we need to combine the coherent QND interaction of Eq.~(\ref{Eq:HamCohAmp}) (QND variables $X_2$ and $P_1$) with the corresponding dissipative interaction; this dissipative interaction
requires the jump operator $\hat{z} = \hat{X}_2 + i \hat{P}_1$, as given in Eq.~(\ref{Eq:zQND}). 

To generate the required dissipation, we again take the engineered reservoir to be a damped auxiliary mode $\hat{c}$ (damping rate $\kappa'$).    Writing this operator in terms of quadratures as $\hat{c} =( \hat{U} + i \hat{V}) / \sqrt{2}$, the required system-bath interaction has the form
\begin{align}\label{Eq:HamDissAmp}
 	\hH_{\rm SB} =   \Lambda \left[ \hat P_1 \hat V + \hat X_2 \hat U  \right].
\end{align}
This interaction preserves the QND structure in the coherent interaction, as it also commutes with $\hat{X}_2$ and $\hat{P}_1$.  One can again confirm that in the Markovian limit of a large $\kappa'$, one recovers the master equation description, with the dissipative rate $\Gamma$ in Eq.~(\ref{Eq.:GeneralMaster}) being given by $ \Gamma = \frac{2 \Lambda^2}{\kappa^{\prime}}$.

The dissipative interaction on its own generates phase-sensitive amplification that can be quantum limited.  Heuristically, this can be understood as arising from a kind of transduction mediated by the reservoir.  
From Eq.~(\ref{Eq:HamDissAmp}) information in the $P_1$ quadrature of cavity 1 drives the auxiliary mode $U$ quadrature.  The $U$ quadrature in turn drives the cavity-2 $P_2$ quadrature, effectively letting $P_1$ drive $P_2$.
As $P_1$ and $X_2$ are QND variables, increasing $\Lambda$ simply increases the gain associated with this process, with no possibility of instability.  An analogous argument of course shows that one obtains reverse gain:  signals incident on $X_2$ will emerge amplified in $X_1$.  Thus, the dissipative amplification here is not directional; directionality is only obtained when this process is matched against its coherent counterpart, as described in Sec.~\ref{Sec.:DPA}. 

The only small non-ideality left in the directional phase-sensitive amplifier of Sec.~\ref{Sec.:DPA} is the presence of reflections at the input
(cf.~Eq.~(\ref{Eq.DirAmpOutputShort})).  Even though there is no gain associated with these, 
one would ideally want them to be exactly zero to protect the signal source.  
We now show that this can be easily accomplished by modifying both the coherent and dissipative interactions used in the scheme, so as to slightly
deviate the QND structure discussed above.  This modification also allows one to cancel reflections of signals and noise at the amplifier output.

To impedance match, we first modify the system-bath Hamiltonian in Eq.~(\ref{Eq:HamDissAmp}) to take the more general form 
\begin{align}\label{Eq.:DPAdissIM}
  \hH_{\rm SB} \equiv&  \hspace{0.4cm}
		        \sqrt{2} \Lambda_U \  \hat U \left( \sin \theta  \hat X_1   + \cos \theta  \hat X_2   \right) \nonumber \\ &
		      + \sqrt{2} \Lambda_V \  \hat V \left( \cos \theta  \hat P_1   + \sin \theta  \hat P_2   \right).
\end{align}
For $\theta = 0$ and $\Lambda_U = \Lambda_V =  \Lambda /\sqrt{2}$ we recover Eq.~(\ref{Eq:HamDissAmp}).  By allowing  
$\Lambda_U \neq \Lambda_V$ , we modify the relative strength of the two QND interactions.  By letting $\theta$ deviate slightly from zero, 
we break the QND structure of Eq.~(\ref{Eq:HamDissAmp}); this will allow us to cancel the unwanted reflections from both cavities. 
In what follows, it will be useful to parametrize the system-bath couplings in terms of a cooperativity $\bar{\mathcal{C}}$ and asymmetry parameter $\alpha$:
\begin{align}
	\bar{\mathcal{C}} & = 4 \Lambda_U \Lambda_V / (\kappa \kappa'), \\
	\alpha &= (\Lambda_V / \Lambda_U)^2,
		\label{Eq:AlphaDefn}
\end{align}

\begin{figure*}
 \centering\includegraphics[width=1.0\textwidth]{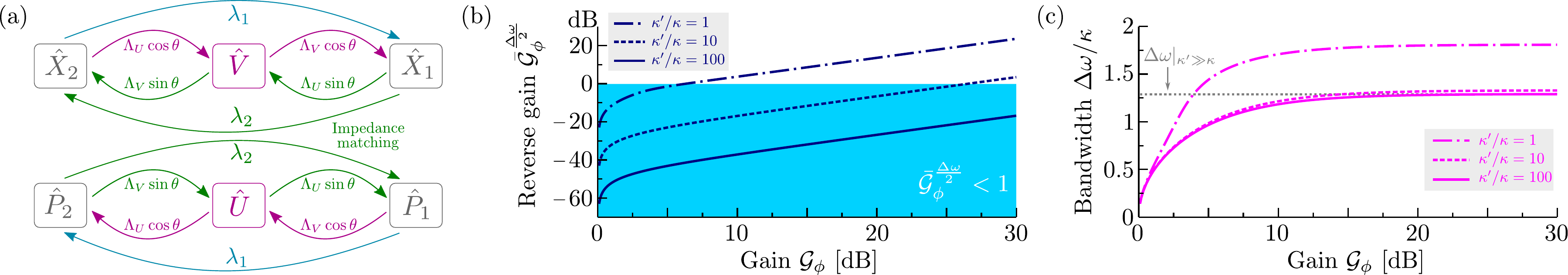} 
	\caption{Properties of the directional phase-sensitive amplifier.  
	(a) Sketch of couplings and drivings used to impedance match the amplifier.   
	(b) Reverse gain $\bar{\mathcal{G}}_{\phi}[\omega]$ for $\omega$ set to  half of the amplification bandwidth $\Delta \omega$, 
	for various choices of the auxiliary-mode damping rate $\kappa'$. 
	On resonance we always have perfect directionality: $\bar{\mathcal{G}}_{\phi}[0]=0$. 
	(c) Amplification bandwidth $\Delta \omega$ as a function of  zero-frequency forward gain $\mathcal{G}_{\phi}$ for various values of $\kappa'$. 
	The amplifier does not suffer from a standard gain-bandwidth constraint.
	\label{Fig:DPADetails}
	}
 \end{figure*}

The general structure of Eq.~(\ref{Eq.:DPAdissIM}) implies that once the auxiliary mode is eliminated, the cavity $X_j$ modes will drive one another in a non-directional way; the same goes for the cavity $P_j$ quadratures.  This is depicted schematically in Fig.~\ref{Fig:DPADetails}(a).  To obtain directionality, we need to cancel the ability of the cavity-$2$ quadratures to drive the corresponding cavity-$1$ quadratures.  We do this in the usual manner:  we balance the dissipative quadrature-quadrature interactions generated by Eq.~(\ref{Eq.:DPAdissIM}) by the coherent versions of these interactions.  
This will require a coherent Hamiltonian of the form:
\begin{align}\label{Eq.:DPAcohIM}
	 \hH_{\rm coh} \equiv& \lambda_1 \hat P_1 \hat X_2  + \lambda_2 \hat P_2 \hat X_1 .  
\end{align}
The second term here is new compared to Eq.~(\ref{Eq:HamCohAmp}), and breaks its QND-structure.  As usual, we balance the above coherent interactions against their dissipative counterparts (as generated by Eq.~(\ref{Eq.:DPAdissIM})) so that the cavity-1 quadratures are not driven by the cavity-$2$ quadratures.  Working through the equations of motion, and focusing on the Markovian limit, we obtain the directionality conditions
\begin{align}
	\label{Eqs:DPALambdaTuning}
 	\lambda_1 =  \kappa   \bar{\mathcal C} \cos^2 \theta,
 \hspace{0.5cm}
 	\lambda_2 = - \kappa \bar{\mathcal C}  \sin^2 \theta.
\end{align}

Using a standard Heisenberg-Langevin analysis, we find the full scattering matrix of the system; tuning the coherent interactions as 
per Eqs.~(\ref{Eqs:DPALambdaTuning}), the scattering will indeed be directional.  Insisting further that
there are no reflections of signals and noise incident on either cavity (i.e.,~impedance matching) leads to an additional condition on the angle $\theta$:
\begin{align}
 	\sin 2 \theta = 1/ \bar{\mathcal C} .
\end{align}
Note that for a large cooperativity $\bar{\mathcal{C}}$, the angle $\theta$ is very close to zero, implying that one is very close to our original
scheme where $\hat P_1$ and $\hat X_2$ are QND variables.

Even after satisfying the above conditions, the cooperativity $\bar{\mathcal{C}}$ and the asymmetry parameter $\alpha$ remain unspecified; they control the final form of the impedance-matched, directional scattering matrix.  Using the above conditions, the full scattering matrix of the system (describing both the principle modes and the auxiliary mode $\hat{c}$) takes a simple form.  Introducing the vector
\begin{align}
\mathbf{W}  =
	\left(
	\begin{array}{cccccc}
 \hat X_{1}
&
  \hat P_{1}
&
  \hat X_{2}
&
  \hat P_{2}
  &
  \hat U
&
  \hat V
	\end{array}
	\right)^T,
\end{align}
the scattering relations at each frequency then take the form $\mathbf{W}_{\rm out} = \tilde{\mathbf{s}} \ \mathbf{W}_{\rm in}$.   At zero frequency, 
the scattering matrix is
\begin{equation}  \label{Eq.DirAmpOutput}
\tilde{\mathbf{s}} =
\left(
\mkern-5mu
\begin{tikzpicture}[baseline=-.65ex]
\matrix[
  matrix of math nodes,
  column sep=-0.5ex,
] (m)
{
 0 & 0 & 0 & 0 & 0 &  - \left[\alpha \mathcal G_{\phi} \right]^{\frac{1}{4}}  
\\[-1.0ex]
 0 & 0 & 0 & 0 & \frac{1 }{ \left[\alpha \mathcal G_{\phi}\right]^{\frac{1}{4}}}  & 0 
\\[-1.0ex]
 \frac{1}{\sqrt{\mathcal G_{\phi}}} & 0 & 0 & 0 & 0 & 0 
\\[-1.0ex]
 0 & \sqrt{\mathcal G_{\phi}} & 0 & 0 & 0 & 0 
\\[-1.0ex]
 0 & 0 & 0 & -  \left[ \frac{\alpha }{\mathcal G_{\phi}  }  \right]^{\frac{1}{4}} & 0 & 0 
\\[-1.0ex]
 0 & 0 &   \left[ \frac{\mathcal G_{\phi}  }{\alpha }   \right]^{\frac{1}{4}} & 0 & 0 & 0 
\\[-1.0ex]
};
\draw[dashed]
  ([xshift=0.5ex]m-1-4.north east) -- ([xshift=0.5ex]m-4-4.south east);
\draw[dashed]
  ([yshift=-0.5ex]m-4-1.south west) -- ([yshift=-0.6ex]m-4-4.south east); 
\end{tikzpicture}
\mkern-5mu
\right) .
\end{equation}
$\mathcal{G}_\phi$ describes the zero-frequency phase-sensitive photon number gain of our amplifier, and it is given by
\begin{equation}
	\sqrt{ \mathcal{G}_\phi}   =  \bar{\mathcal{C}} \left( 1 + \sqrt{1 - \frac{1}{\bar{\mathcal{C}}} } \right).
\end{equation}

The upper $4 \times 4$ block describes an ideal, directional phase-preserving amplifier.  As for the full $6 \times 6$ scattering matrix, 
it describes a kind of ``squeezing circulator", where the input on port $j$ emerges from port $j+1$ after having undergone a squeezing transformation.
Crucially,  the squeezing parameters or gains for each of these transformations are not all equal.

While a squeezing circulator may have interesting applications, if the goal is amplification, it represents a potential hazard.  As indicated by 
Eq.~(\ref{Eq.DirAmpOutput}), incident noise from the auxiliary mode will emerge from the output of cavity $1$, having undergone a squeezing transformation with gain $\sqrt{\alpha \mathcal{G}_{\phi}}$.  To protect the signal source at the amplifier input, we do not want to amplify any fluctuations emerging from the auxiliary mode.  Hence, the ideal choice is to null this effective gain by tuning the asymmetry $\alpha$ to satisfy
\begin{equation}
	 \alpha  =  1/\mathcal{G}_\phi,
	 \label{Eq:alphaOpt}
\end{equation}
Tuning the asymmetry of the couplings to the auxiliary cavity in this manner ensures that one can have large, directional gain, without unduly large amounts of noise emerging from the amplifier input port.

The presented phase-sensitive amplifier has several highly desirable properties: it is quantum limited, directional and has no gain-bandwidth limitation. 
However, an experimental implementation in a superconducting circuit setting will also face some technical challenges.  Most notably, a  
straightforward implementation requires 6 pump tones to be applied with excellent control over their amplitudes and relative phases.  
While demanding, experiments with analogous levels of complexity and multiple pumps have recently been performed in circuit QED architectures, see
e.g.,~Refs.~\onlinecite{Sliwa2015,Shankar2013}.


\subsubsection{Frequency dependence}

The full scattering matrix can also be easily calculated at non-zero frequencies.  The relevant forward gain of the amplifier describes signals incident in the $P$ quadrature of cavity 1 emerging in the $P$ quadrature of the output from cavity 2.  Assuming that we chose parameters to impedance match (as described), and that we further tune the asymmetry parameter $\alpha$ to minimize noise as per Eq.~(\ref{Eq:alphaOpt}), the forward 
photon number gain is given by:
\begin{align}
 	\mathcal G_{\phi} [\omega]  \equiv \left| \tilde{s}_{42} \right|^2       =&  
			\frac{\mathcal G_{\phi}     \left(1 +\frac{\omega^2}{\kappa^{\prime 2}}\right)}
			     {\left(1+\frac{\omega ^2}{\kappa ^2}\right)^2  
				+ \frac{\omega ^2}{\kappa^{\prime 2}} \left(1+ \frac{4 \omega ^4}{\kappa ^4}\right)
				-   \frac{ 4\omega ^4}{\kappa ^3 \kappa^{\prime  }}  }. 
\end{align}
As already discussed, the zero-frequency gain $\mathcal{G}_\phi$ can be made arbitrarily large by simply increasing the various couplings 
(i.e.,~$\bar{\mathcal{C}}$); the linear system never exhibits any instability. 
In the Markovian limit $\kappa' \gg \omega$, the frequency dependence of the gain is extremely simple:  it is simply a Lorentzian squared, with a bandwidth $\Delta \omega \sim \kappa$ which is {\it independent} of the zero-frequency gain.  Thus, this amplifier is not constrained by any fundamental gain-bandwidth limitation. 

Including non-Markovian effects (i.e.,~finite $\omega / \kappa^{\prime}$), the frequency dependence is slightly more complex, but the ultimate bandwidth is still set by $\kappa$, irrespective of the size of the zero-frequency gain.

While deviations from the Markovian limit do not degrade amplification, they impact the directionality of the amplifier.  In the ideal Markovian limit, signals incident on cavity 2 in either quadrature never emerge from cavity 1.  For finite $\omega / \kappa'$, this is no longer true:  now, 
the reverse-gain scattering matrix element $\tilde{s}_{13}$ becomes non-zero, implying that incident signals on $\hat X_2$ can emerge from $\hat X_1$. 
We find:
\begin{align}
 	\bar{ \mathcal G}_{\phi} [\omega] \equiv \left|\tilde{s}_{13} \right|^2  =& 
			\frac{\mathcal G_{\phi}     \left( \frac{\omega^2}{\kappa^{\prime 2}}\right)}
			     {\left(1+\frac{\omega ^2}{\kappa ^2}\right)^2  
				+ \frac{\omega ^2}{\kappa^{\prime 2}} \left(1+ \frac{4 \omega ^4}{\kappa ^4}\right)
				-   \frac{ 4\omega ^4}{\kappa ^3 \kappa^{\prime  }}  } . 
\end{align}
As expected, the optimal situation is clearly in the Markovian limit where $\kappa^{\prime} \gg \kappa$.  In this limit, one has purely directional amplification over the full bandwidth $\kappa$ of the principle cavity modes.


\subsubsection{Added noise for finite frequency}

An ideal phase-preserving amplifier can amplify a single quadrature without any added noise \cite{Caves1982}.
From the scattering relations of Eq.~(\ref{Eq.DirAmpOutput}) it immediately follows that our scheme reaches this quantum limit on resonance. 
For completeness, we also present the added noise at finite frequency, again focusing on the impedance-matched version of the amplifier.
We calculate the added noise of our amplifier in the standard
manner \cite{Clerk2010}, by calculating the noise in the cavity-2 output $P$ quadrature (symmetrized spectral density $\bar S_{P_2}[\omega]$), and then referring this back to the input.  Expressing this added noise as an effective number of thermal quanta, 
and focusing on the Markovian limit $\kappa^{\prime} \gg \kappa $, we find

\begin{align}
 \bar n_{P_2, \rm add}[\omega] 
		=&   \frac{\omega ^2}{\kappa ^2 } \left(
		  \frac{  \bar n_c^T + \frac{1}{2}}{ \sqrt{ \mathcal{G}_{\phi} \alpha}}  
		  +   \left[ 1 + \frac{\omega ^2}{\kappa ^2 }    \right]   
		   \frac{\left(\bar n_2^T+\frac{1}{2}\right) }{\mathcal G_{\phi}} \right)
			+ \mathcal O \left[ \frac{\omega}{\kappa^{\prime }}\right].
\end{align}
Note that we have left the asymmetry parameter $\alpha$ (cf.~Eq.~(\ref{Eq:AlphaDefn})) unspecified here.  The added noise always vanishes at $\omega=0$, irrespective of the gain.  Furthermore, for a fixed value of $\alpha$ the added noise vanishes at all frequencies in the large gain limit, implying that one is quantum limited at all frequencies.  If however one tunes $\alpha = 1 / \mathcal{G}_\phi$ to minimize the noise hitting the input port, then the added noise is non-zero at finite frequencies even in the large gain limit.


\section{Conclusion}

We have presented an extremely general yet simple method for achieving directional behavior in coupled photonic systems, based on matching a
given (reciprocal) coherent interaction with the corresponding dissipative version of the interaction.  We demonstrated how this principle could be used to construct both isolators and directional, quantum-limited amplifiers.  In particular, our approach allows the construction of a directional phase-sensitive amplifier that is not limited by a standard gain-bandwidth constraint.  The recipe we present is not tied to a particular realization, and could be implemented in photonic systems, microwave superconducting circuits, and optomechanical systems.

Finally, while our focus here has been on bilinear interactions between two principle cavity modes, a similar approach of balancing coherent and dissipative interactions could be used to make nonlinear interactions directional, and could be used in more complex cavity lattice structures.  Understanding how this form of reciprocity breaking leads to useful functionalities and possibly new photonic states in such systems will be the subject of future work.


\acknowledgments{
We thank Joe Aumentado, Michel Devoret, Archana Kamal and Leonardo Ranzani for useful conversations.
This work was supported by the DARPA ORCHID program through a grant from AFOSR.
}


\appendix 

\section{General approach to directionality}\label{AppendixA}

We show here that our basic recipe of balancing coherent and dissipative interactions can make any factorizable interaction between two quantum systems
directional. Consider two bosonic systems $1$ and $2$, and consider a general interaction Hamiltonian of the form:
\begin{align}
 \hH_{\rm coh}  =& \frac{\lambda}{2} \left(  \hat o_1 \hat o_2 + \hat o_2^{\dag} \hat o_1^{\dag} \right).
\end{align}
Here, $\hat{o}_j$ is a system $j$ operator ($j=1,2$), implying that
\begin{align}
  \left[\hat o_1^{(\dag)}, \hat o_2^{(\dag)}  \right] = \left[\hat o_1, \hat o_2^{\dag}  \right]  = \left[\hat o_1^{ \dag }, \hat o_2  \right]= 0 .
\end{align}
The operators $\hat{o}_j$ are otherwise arbitrary; in the case where our systems are cavity modes, they could be nonlinear combinations of creation/destruction operators, and/or non-Hermitian.

To make the above general interaction directional, we need to introduce its dissipative counterpart.  We do this by coupling to a suitably engineered reservoir which couples to both subsystems, and which gives rise to a Lindblad master equation of the form
\begin{align}
 \frac{d}{dt} \hat \rho = & - i \left[\hH_{\rm coh}, \hat \rho \right] + \Gamma \mathcal L \left[\hat o_1 +e^{i\varphi} \hat o_2^{\dag} \right] \hat \rho.
\end{align}
We have a single dissipator with jump operator $\hat{z} = \hat o_1 +e^{i\varphi} \hat o_2^{\dag}$, with a dissipative rate $\Gamma$.
The phase $\varphi$ appearing in $\hat{z}$ will be kept general for the moment; we see that by tuning it correctly, we obtain the desired directionality.

It is now straightforward to calculate the equations of motion for some arbitrary operators $\hat A_n,( n \in 1,2)$ for each subsystem; 
as usual, all system-$1$ operators commute with all system-$2$ operators.  
Using the above Lindblad master equation we obtain
\begin{align}
 \frac{d}{dt} \ev{\hat A_1} = & 
- \frac{i}{2}     \left[   \lambda +    \Gamma e^{- i \widetilde \varphi }  \right]  \ev{ \left[ \hat A_1 ,\hat o_1^{\PD}  \right]  
				\hat o_2        \hat \rho  }
\nonumber \\ &
- \frac{i}{2}     \left[   \lambda + \Gamma e^{  i \widetilde \varphi }   \right]  \ev{ \left[ \hat A_1 ,\hat o_1^{\dag}\right]  
				\hat o_2^{\dag} \hat \rho  }
		+ \Gamma \ev{ \hat A_1 \mathcal L[\hat o_1] \hat \rho} ,
\nonumber \\  
 \frac{d}{dt} \ev{\hat A_2} = &    
- \frac{i}{2} 
			   \left[ \lambda  -  \Gamma e^{- i \widetilde \varphi }   \right] 
				\ev{  \left[ \hat A_2,  \hat o_2^{\PD}   \right]   \hat o_1   \hat \rho } 
\nonumber \\ &
- \frac{i}{2} 
			  \left[ \lambda  - \Gamma e^{i\widetilde \varphi  }  \right] 
			\ev{  \left[ \hat A_2,  \hat o_2^{\dag}\right]   \hat o_1^{\dag} \hat \rho }
		+ \Gamma \ev{ \hat A_2 \mathcal L[\hat o_2^{\dag}] \hat \rho} ,
\end{align}
with $\widetilde \varphi = \varphi - \frac{\pi}{2}$.
The first two terms on the RHS of each equation describe the effects of interactions between the two systems (both dissipative and coherent), 
while the respective third term describes a local, generalized damping induced by the engineered reservoir.
To obtain directionality, we take $\varphi = \frac{\pi}{2}$.  Focusing only on the terms coupling the two cavities, the EOM take the form: 
\begin{align} \label{Eq.AppGeneralEOM}
 \frac{d}{dt} \ev{\hat A_1} =& 
 			- \frac{i}{2} \left[ \lambda  +  \Gamma  \right] 
 			\left\{
 			  \ev{ \left[ \hat A_1 ,\hat o_1^{\PD}      \right]  \hat o_2        \hat \rho  }
 			+ \ev{ \left[ \hat A_1 ,\hat o_1^{\dag}\right]  \hat o_2^{\dag} \hat \rho  }
 			\right\} , 
\nonumber \\  
 \frac{d}{dt} \ev{\hat A_2} =&    
 			- \frac{i}{2} \left[ \lambda  -  \Gamma  \right] 
 			\left\{  \ev{  \left[ \hat A_2,  \hat o_2^{\PD}       \right]   \hat o_1        \hat \rho } 
 			+        \ev{  \left[ \hat A_2,  \hat o_2^{\dag}\right]   \hat o_1^{\dag} \hat \rho } \right\},
\end{align}
where again, we have dropped the local generalized damping terms.
This form makes it obvious that for $\Gamma = \lambda$ we decouple system $2$ from system $1$:  system-$2$ observables are not influenced
at all by system $1$, whereas system $1$ is influenced by system $2$.  We thus have made the original, general interaction described by $\hH_{\rm coh}$ directional, by picking a suitable form for the dissipative jump operator and for the corresponding dissipative rate.  This simply corresponds to our general recipe of balancing a given coherent interaction against its dissipative counterpart.  Note that if we had instead made the choice $\varphi = - \pi/2$, the direction of the final nonreciprocal interactions would be flipped, with cavity $1$ now influencing cavity $2$.

As an example we consider the directional DPA discussed in Sec.~\ref{Sec.:DPA}. 
There we would have the correspondence $\hat o_1 = \hat X_2$ and $\hat o_2 = \hat P_1$.
Taking these together with the equations for the expectation values in Eq.~(\ref{Eq.AppGeneralEOM}), we recover that the
QND observables $\hat X_2$ and $\hat P_1$ are not affected by the interaction, but the expectation values for the remaining quadratures, i.e., setting $\hat A_1 = \hat P_2$ and $\hat A_2 = \hat X_1$, become   
\begin{align}
 \frac{d}{dt} \ev{\hat P_2} = & 
 			-  \left[ \lambda  +  \Gamma  \right] \ev{   \hat P_1      },  
\hspace{0.15cm} 
 \frac{d}{dt} \ev{\hat X_1} =      
 			+   \left[ \lambda  -  \Gamma  \right]  \ev{ \hat X_2 }  ,
\end{align}
which coincides with our former result, cf.~Eq.~(\ref{Eq.:EoMexpectDPA}).

Finally, the above construction also applies directly to fermionic systems, if one takes the operators $\hat{o}_j$, $\hat{A}_j$ to be even in creation/destruction operators (implying that cavity-$1$ and cavity-$2$ operators commute with one another). 

\section{Waveguide as an engineered reservoir}\label{AppendixB}

We show here how a simple 1D transmission line or waveguide can be used as the engineered reservoir needed in the dissipative isolator scheme of Sec.~\ref{subsec:IsolatorIntro} 
(i.e.,~it generates the dissipator in the master equation of Eq.~(\ref{Eq.:GeneralMaster}) with jump operator $\hat{z} = \hat{d}_1 + \hat{d}_2$).  We take a standard approach for the case where the waveguide dispersion is linear over all frequencies of interest (see, e.g.,~Appendix A of Ref.~\cite{Chang2011}).  Working in an interaction picture at the frequency $\omega_{\rm cav}$ of the two principle cavities, the Hamiltonian of the waveguide takes the form
\begin{equation}
	\hH_{\rm W} =
		 \hbar \vwg  \int dx
		\left(
			\hat{c}^\dagger_R (- i \partial_x - k_0) \hat{c}_R + \hat{c}^\dagger_L (i \partial_x - k_0) \hat{c}_L
		\right),
\end{equation}
where $k_0 = \omega_{\rm cav} / \vwg$, and we have omitted the explicit $x$ dependence of the waveguide fields $\hat{c}_R(x)$, $\hat{c}_L(x)$.  

Using the total Hamiltonian $\hH = \hH_{\rm W} + \hH_{\rm SB}$ (where $\hH_{\rm SB}$ is given in Eq.~(\ref{Eq:SysBathWG})), we find the equation of motion
\begin{align}
	\left(  \frac{1}{\vwg} \partial_t + \partial_x \right) \hat{c}_{R}(x,t) & = 
		i k_0 \hat{c}_R(x,t)  
		\nonumber \\
		&	+ i \sum_{j=1,2}
				\frac{\Gamma}{2 \vwg}
		 		\delta(x- x_j) \hat{d}_j, 
\end{align}
with an analogous equation for $\hat{c}_{L}(x,t)$.  The delta-function source term leads to a discontinuity in each waveguide field, e.g.,
\begin{align}
	c_{R}(x_j + \eta,t) & = 
		c_{R}(x_j - \eta,t) + i \frac{\Gamma}{2 \vwg} \hat{d}_j(t).  
\end{align}
Introducing input and output $R$ fields associated with cavity $j$ in the natural manner, this takes the form of a standard input-output relation:
\begin{align}
	c_{R, {\rm out}}(x_j,t) & = 
		c_{R, {\rm in}}(x_j,t) + i \frac{\Gamma}{2 \vwg} \hat{d}_j(t).
\end{align}
A similar equation (and definition of input and output fields) holds for the $L$ field.

Using the fact that fields propagate freely between the cavities and $l \equiv x_2 - x_1>0$, we have
\begin{align}
	\hat{c}_{R,{\rm in}}[x_2,\omega] &=		e^{i k[\omega] l } \hat{c}_{R,{\rm out}}[x_1,\omega], \\
	\hat{c}_{L,{\rm in}}[x_1,\omega] &=		e^{i k[\omega] l } \hat{c}_{L,{\rm out}}[x_2,\omega], 
\end{align}
where we have Fourier transformed in time and defined $k[\omega] = k_0 + \omega / \vwg$.  We can finally substitute these results into the Heisenberg equations of motion for the cavity operators
$\hat{d}_j$.  Fourier transforming, they take the form
\begin{align}
	\label{Eq.:WaveguideEOM}
	\left(	
		\begin{array}{cc}
  			-i \omega +  \frac{\Gamma}{2} 	&	e^{i k[\omega] l} \frac{\Gamma}{2}  
				\\[2mm]
  			e^{i k[\omega] l} \frac{\Gamma}{2}  		&	-i \omega +  \frac{\Gamma}{2}
		\end{array}
	\right)
	\left( \begin{array}{c}
  			\hat{d}_1[\omega]  \\[2mm]
			\hat{d}_2[\omega] 
	\end{array} \right)
		& =
		i \sqrt{\frac{\Gamma}{2}} 
		\left( \begin{array}{c}
  			\hat{\xi}_1[\omega]  \\[2mm]
			\hat{\xi}_2[\omega] 
	\end{array} \right),
 \end{align}
 where the noise operators $\xi_j$ are
 \begin{align}
 	\xi_1[\omega] &=
		 \sqrt{\vwg} \left(
			\hat{c}_{R,{\rm in}}[x_1,\omega] 
			+ e^{i k[\omega] l} \hat{c}_{L,{\rm in}}[x_2,\omega] 
	 \right), \\
 	\xi_2[\omega] &=
		 \sqrt{\vwg} \left(
			e^{i k[\omega] l} \hat{c}_{R,{\rm in}}[x_1,\omega] +  \hat{c}_{L,{\rm in}}[x_2,\omega]  \right). 
\end{align}
Note that we have only retained terms associated with the coupling to the waveguide, as our goal here is to see the form of the dynamics it induces for the cavities.

Consider the case where for all frequencies of $\omega$ of interest, $\omega \ll  1/\tau$, where $\tau = l/ \vwg$ is the propagation time between the cavities.  
In this case, we can omit the effects of non-zero $\tau$ in Eqs.~(\ref{Eq.:WaveguideEOM}), and replace $k[\omega] $ by $k_0$.  
Note that in general, we are only concerned with the cavity dynamics on frequencies that are most comparable to $\kappa$, hence this approximation requires
$\kappa \tau \ll 1$.  Once we make this Markovian approximation, Eqs.~(\ref{Eq.:WaveguideEOM}) become local in time, and have the same structure as 
Eq.~(\ref{Eq.:EoMexpIso}) and Eq.~(\ref{Eqs:AuxCavEOM}) in the main text.  Comparing the form of the equations, we see that 
the cavity induces both a coherent hopping interaction between the cavities (amplitude $J_{\rm ind} = \Gamma \sin k_0 l / 2$) and an induced dissipative hopping interaction (strength $\Gamma_{\rm ind} = 
\Gamma \cos k_0 l $).  

The induced coherent interaction here always corresponds to a real hopping $J_{\rm ind}$.  As such, we cannot use the waveguide to provide both the interactions needed for our isolator scheme (i.e.,~one cannot satisfy the directionality condition of Eq.~(\ref{Eq:DirectionalHoppingCond}) using the waveguide alone).  Instead, we can use the waveguide solely to provide the dissipative interaction needed for the scheme.  We thus require the distance between the cavities to be chosen such that there is no induced coherent hopping interaction, i.e.,
\begin{equation}
	k_0 l = n \pi, n \in \mathbb{Z}.
	\label{Eq:WGDistanceCond}
\end{equation}

If  the integer $n$ in Eq.~(\ref{Eq:WGDistanceCond}) is even, then the noise operators in Eqs.~(\ref{Eq.:WaveguideEOM}) are identical:  $\hat{\xi}_1 = \hat{\xi}_2$.  For this choice (and in the Markovian limit), the dissipative interactions induced by the waveguide are completely described by the dissipator $\mathcal{L}[z]$ in Eq.~(\ref{Eq.:GeneralMaster}) with the choice $\hat{z} = \hat{d}_1 + \hat{d}_2$ (as can be shown using standard techniques \cite{Gardiner2004}).  The dissipative interactions generated by the waveguide in this limit are thus also equivalent to those generated by the auxiliary cavity implementation (cf.~Eq.~(\ref{Eq:SysBath})) discussed in Sec.~\ref{subsubsec:AuxMode}.  For $n$ odd in Eq.~(\ref{Eq:WGDistanceCond}), one generates the dissipator $\mathcal{L}[z]$ with $\hat{z} = \hat{d}_1 - \hat{d}_2$; upon making a gauge change $\hat{d}_2 \rightarrow - \hat{d}_2$, this is of course equivalent to having $\hat{z} = \hat{d}_1 + \hat{d}_2$.


\end{document}